\documentclass[
aps,
pra,
amsmath,amssymb,
showpacs,
preprint,
reprint,
]{revtex4-2}

\usepackage[linkcolor=red, citecolor=blue, colorlinks=true, urlcolor=blue]{hyperref}
\usepackage{graphicx}
\usepackage{dcolumn}
\usepackage{bm}
\renewenvironment{widetext@grid}{%
  \par\ignorespaces
  \setbox\widetext@top\vbox{%
   \vskip15\p@
   \hb@xt@\hsize{%
    \leaders\hrule\hfil
    \vrule\@height6\p@
   }%
   \vskip6\p@
  }%
  \setbox\widetext@bot\hb@xt@\hsize{%
    \vrule\@depth6\p@
    \leaders\hrule\hfil
  }%
  \onecolumngrid
  \let\set@footnotewidth\set@footnotewidth@ii
}{%
  \par
  \twocolumngrid\global\@ignoretrue
  \@endpetrue
}%

\makeatother

\begin{document}

\title[]{Accessing  Kardar-Parisi-Zhang universality sub-classes with exciton polaritons}

\author{Konstantinos Deligiannis$^{1}$, Davide Squizzato$^{2, 3}$, Anna Minguzzi$^1$, L\'eonie Canet$^{1, 4}$}
\affiliation{$^1$Universit\'e Grenoble Alpes, CNRS, LPMMC, Grenoble, France, \\
$^2$Dipartimento di Fisica, Universit\`{a} La Sapienza, Rome, Italy, 00185,\\ $^3$Istituto Sistemi Complessi, Consiglio Nazionale delle Ricerche, Universit\`{a} La Sapienza Rome, Italy, 00185, \\
$^4$Institut Universitaire de France, Paris, France, 75000}

\date{\today}

\begin{abstract}
Exciton-polariton condensates under driven-dissipative conditions are predicted to belong to the Kardar-Parisi-Zhang (KPZ) universality class, the dynamics of the condensate phase satisfying the same equation as for classical stochastic interface growth at long distance. We show that by engineering an external confinement for one-dimensional polaritons we can access two different universality sub-classes, which are associated to the flat or curved geometry for the interface. Our results for the condensate phase distribution and correlations match with great accuracy with the exact theoretical results for KPZ: the Tracy-Widom distributions (GOE and GUE) for the one-point statistics, and covariance of Airy processes (Airy$_1$ and Airy$_2$) for the two-point statistics. This study promotes the exciton-polariton system as a compelling platform to investigate KPZ universal properties.
\end{abstract}

\maketitle

\section{Introduction}
Phase transitions have been at the heart of statistical physics over the past 60 years, and a central issue for most areas of physics. Whereas a thorough understanding of critical behaviours has been acquired for equilibrium systems, the theoretical description of non-equilibrium phase transitions, in particular the ones involving non-equilibrium steady states, is still a major challenge and has been the subject of intense work in the last decades. Remarkably, self-organised criticality can emerge in non-equilibrium systems, leading to the onset of scale invariance without the need to tune any external parameter. This is realised in the celebrated Kardar-Parisi-Zhang (KPZ) equation \cite{KardarParisiZhang}. Whereas it was originally derived to describe kinetic roughening of interfaces undergoing stochastic growth \cite{HalpinHealyZhang}, the KPZ critical properties have been shown to arise in many non-equilibrium or disordered systems, ranging from turbulent liquid crystals \cite{TakeuchiSano1} to non-equilibrium hydrodynamics \cite{Spohn2016} to name a few.
 
{More recently, Bose-Einstein condensates of exciton-polaritons (EP) \cite{EPBEC}, a  quantum fluid with markedly different properties from equilibrium Bose-Einstein condensate of ultracold atoms, have proven to be a promising playground to observe KPZ universal properties. EP are bosonic quasi-particles arising from the strong coupling of photons to excitons (electron-hole bound states) realised in a  semiconductor microcavity. They are formed under intrinsically driven-dissipative conditions, since one has to introduce an optical pump to overcome the leakage of photons out of the cavity mirrors and maintain a steady state. Properties of EP  have been thoroughly investigated both experimentally and theoretically \cite{CarusottoCiuti}. Recently, a striking connection to KPZ universality has been brought out by several theoretical approaches \cite{Altman2015,Ji2015,GladilinJiWouters}. More precisely, the dynamics of the phase of the condensate wavefunction at long distances has been shown to obey the KPZ equation, and KPZ scaling has been reported in various conditions \cite{Zamora2017,He2017,Comaron2018}. In particular, the KPZ exponents were found in numerical simulations of the one-dimensional \cite{HeSiebererAltmanDiehl} and two-dimensional \cite{Mei2020} EP systems, as well as of photonic cavity arrays \cite{Amelio2019}.}

However, the KPZ universality class encompasses much more than mere scaling. In particular, the exact long time probability distribution of the fluctuations of the height has been determined for a number of systems with a one-dimensional growing interface \cite{SasamotoSpohn, Kriecherbauer, Corwin}. A remarkable feature is that, while these systems share the same critical exponents, their probability distribution depends on the initial conditions of the growth, thereby distinguishing three main geometry-dependent universality sub-classes. For an initially flat, respectively curved, interface, the probability distribution coincides with that of the largest eigenvalue of random matrices in the  Gaussian Orthogonal Ensemble (GOE) \cite{CalabreseLeDoussal, LeDoussal, BaikRains2, BaikRainsbook, Sasamoto, FerrariSpohn}, respectively Gaussian Unitary Ensemble (GUE) \cite{AmirCorwinQuastel, Calabrese, SasamotoSpohn2, BaikDeiftJohansson, Johansson}, unveiling a non-trivial connection with random matrix theory,  where these distributions, called Tracy-Widom (TW), originally emerged \cite{TracyWidom}. The third sub-class corresponds to Brownian, also called stationary, initial conditions, with fluctuations following a Baik-Rains (BR) distribution \cite{ImamuraSasamoto1, ImamuraSasamoto2}.
    
These sub-classes also differ at the level of two-point statistics. More specifically, it was shown that the spatial correlations of the height fluctuations of the one-dimensional growing interface are identical to those of stochastic processes called Airy$_1$ \cite{BorodinFerrariSasamoto, Sasamoto} and Airy$_2$ \cite{PrahoferSpohn_2pt1, ProlhacSpohn} for the flat and curved interface respectively. On the experimental side, the realisation of growing interfaces in turbulent liquid-crystal systems stands as the most advanced platform to study one-dimensional KPZ dynamics. In these experiments, both the one-point and two-point statistics have been measured for both the flat and curved geometries, and they confirm the theoretical results with impressive accuracy \cite{TakeuchiSano1, TakeuchiSano2, TakeuchiSanoSasamotoSpohn}. 
     
In this work, we show that EP condensates appear in many respects as a very versatile set-up to futher investigate KPZ dynamics. While some of the advanced KPZ features were already observed in numerical simulations of EP condensates \cite{Davide}, in particular the TW-GOE distribution for the flat geometry, as well as the BR distribution for the stationary (Brownian) case, in the present paper we demonstrate that the curved KPZ sub-class can also be realised by tailoring a confinement potential that effectively bends the phase profile. Using numerical simulations we find that in the presence of this confinement the phase fluctuations follow the expected TW-GUE distribution. Moreover, we provide the first study of the two-point statistics of the phase of the EP. We first determine the scaling function, which displays similar features for all sub-classes. We then compute the two-point spatial correlations of the fluctuations of the phase in both geometries, and show that they reproduce with great accuracy the expected theoretical ones related to the Airy$_1$ and Airy$_2$ processes, although only locally for the curved case since the condensate phase is bent only over a limited space region. Our study hence shows that all the geometrical KPZ sub-classes can be accessed in EP condensates.

\section{Model for the dynamics of the EP condensate}
Our starting point is the mean-field generalised Gross-Pitaevskii equation for the EP Bose-Einstein condensate under incoherent pumping formulated in \cite{WoutersCarusotto},
\begin{align}
i \hbar\partial_t \psi = \left[ \mathcal{F}^{-1}[E_{LP}(k)]+\frac{i\hbar}{2}\left( Rn_r - \mathcal{F}^{-1}[\gamma_l(k)] \right) \nonumber \right.\\ \left.
+ \hbar g_{\textrm{int}}\left\lvert \psi \right\rvert^2 \right] \psi \,,
\label{eq: model}
\end{align}
where $\psi$ is the condensate wavefunction, $\mathcal{F}^{-1}[E_{LP}(k)]$ denotes the inverse Fourier transform of the dispersion relation of the lower-polariton branch in momentum space, $Rn_r$ is the amplification term, $\gamma_l(k)$ is the loss rate of polaritons, $g_{\textrm{int}}$ is the polariton-polariton interaction strength and $n_r$ is the reservoir density, whose evolution obeys the phenomenological rate equation $\partial_t n_r=P-\gamma_r n_r -R n_r |\psi|^2$,  where $P$ is the pumping strength and $\gamma_r$ the reservoir loss rate. Under the assumption that the time scales of the reservoir and of the condensate are well separated \footnote{Note that one can derive the mapping to the KPZ equation directly from the two-equation system without assuming the separation of time scales of the condensate and reservoir dynamics \cite{DavideKonstantinosLeonieAnna}.}, one may integrate out the reservoir dynamics, which, for sufficiently small field amplitudes, leads to a stochastic equation analogous to a complex Ginzburg-Landau equation,
\begin{align}
i \hbar \partial_t \psi = & \left[\mathcal{F}^{-1}[E_{LP}(k)]  +V(x) + \hbar g_{\text{int}}\left\lvert \psi\right\rvert^2 \right] \psi \nonumber \\
& + \frac{i\hbar}{2}\left[\frac{PR}{\gamma_r}-\frac{PR^2}{\gamma_r^2} \left\lvert \psi \right\rvert^2 - \mathcal{F}^{-1}[{\gamma_l}(k)] \right] \psi + \hbar\xi \,,
\label{eq: gGPEplus}
\end{align}
where the noise $\xi$, which arises from both dissipation and pumping stochastic processes, is complex and has zero mean and covariance $\left\langle \xi(x,t) \xi^*(x', t') \right\rangle = 2\sigma\delta(x-x') \delta(t-t')$ with $\sigma = \gamma_{l,0}(p+1)/2$ \cite{SiebererBuchholdDiehl2016}. This equation further accounts for two effects which were shown to be important to relate the KPZ regime to actual experimental systems: the quartic correction to the dispersion relation $E_{LP}(k)=\hbar \omega_{0,LP}+\frac{\hbar^2}{2m}k^2 - \frac{1}{2 \hbar \Omega} \left(\frac{\hbar^2}{2m}\right)^2 k^4$  with $\Omega$ the Rabi frequency and a momentum-dependent loss rate of polaritons ${\gamma_l}(k)=\gamma_{l,0} + k^2 \gamma_{l,2}$. We have also introduced a confinement potential $V(x)$, which is the cornerstone of this work.

\section{KPZ mapping with the confinement}
To establish the mapping to the KPZ equation, one usually decomposes the wavefunction $\psi$ in the density-phase representation $\psi=\sqrt{\rho} e^{i \theta}$ and expands Eq.~(\ref{eq: gGPEplus}) for small variations around the mean-field solution and in powers of gradients. Following a similar strategy, one finds that in the presence of the static potential $V(x)$, the dynamics of the phase of the EP condensate at long distances obeys an inhomogeneous KPZ equation (see  Sec. \ref{S2} for the derivation),
\begin{equation}
\partial_t \theta = \nu(x) \partial_x^2 \theta + \dfrac{\lambda(x)}{2} \left( \partial_x \theta \right)^2 + \sqrt{D(x)}\eta \,,
\label{eq: KPZ}
\end{equation}
where $\eta$ is a white noise with zero mean and covariance $\left\langle \eta(x,t) \eta(x', t') \right\rangle = 2\delta(x-x') \delta(t-t')$, and 
\begin{align}
& \nu(x)= \frac{\gamma_{l,2}}{2} + \frac{\hbar \rho_0(x)}{m}\tilde{u}(x) \,,\nonumber\\
& \lambda(x)=-\frac{\hbar}{m}+2\gamma_{l,2}\rho_0(x)\tilde{u}(x)\,,\nonumber\\ 
& D(x)=\frac{\sigma}{2\rho_0(x)}(1+4\rho_0^2(x)\tilde{u}^2(x))\,,\nonumber\\
& \tilde{u}(x)=\dfrac{\frac{\hbar}{2m} \left[ \frac{(\partial_x \rho_0)^2}{2\rho_0^3} - \frac{\partial_x^2 \rho_0}{2\rho_0^2} \right]- \frac{\gamma_{l,2}}{2} \frac{(\partial_x \rho_0)( \partial_x \theta_0)}{\rho_0^2} - g_{\text{int}}}{\frac{\gamma_{l,2}}{2}\left[ \frac{(\partial_x \rho_0)^2}{2\rho_0^2} - 2(\partial_x \theta_0)^2 \right] - \frac{\hbar}{m}\partial_x^2 \theta_0 + \tilde{p}(x)}\,, \label{eq:param}
\end{align}
where $\theta_0(x,t)$ and $\rho_0(x)$ are the mean-field inhomogeneous solutions and $\tilde{p}(x)=\left(p-1-\frac{2pR}{\gamma_r}\rho_0(x)\right)\gamma_{l,0}$ with dimensionless pumping parameter $p=PR/(\gamma_{l,0} \gamma_r)=P/P_{\textrm{th}}$, $P_{\textrm{th}}$ being the threshold pumping strength for condensation.

When $V(x)=0$, one recovers the standard mapping to the homogeneous KPZ equation where the parameters $\nu$, $\lambda$ and $D$ are constant \cite{Davide}. With a non-vanishing potential, these parameters continuously vary with space. One can get an intuition of how this may affect the dynamics by noting that the average velocity of the phase is proportional to the KPZ non-linearity $\lambda$. Qualitatively, when $V\neq 0$, the non-linearity $\lambda(x)$, and thus the velocity, increases where the potential is larger. Therefore, if one implements a suitable enhancement of the potential at the boundaries, one may induce an effective drag at the boundaries prone to bend the phase profile. We found that this can indeed be realised.

In our work, we consider a harmonic potential 
\begin{equation}
V(x) = \frac{1}{2}m \omega_0^2 x^2,
\label{eq:pot}
\end{equation}
where the frequency trap $\omega_0$ can be adjusted. We found that the most favorable trap to study KPZ properties is a shallow parabola (we choose $\omega_0 \simeq 4\times10^{-4} \gamma_{l,0}$ in the following), since it allows one to keep the density fluctuations tame and the KPZ parameters slowly varying in the vicinity of $x=0$. Furthermore, we checked that the  curved KPZ universality sub-class is robust and can be observed for different confinement potentials, in particular Gaussian walls (see Sec. \ref{S4}).

\section{Numerical simulations}
We solve the generalised Gross-Pitaevskii equation (\ref{eq: gGPEplus}) for a  system size of $L/\hat{x}=2^{10}$, where we have chosen $\hat{x}=\sqrt{\hbar/(2m\gamma_{l,0})}\simeq \,2\mu$m, $\hat{t} = \gamma_{\ell,0}^{-1}$ and $\hat{\epsilon}=\hbar \gamma_{l,0}$ as units for length, time and energy. We take for the parameters of Eq.~(\ref{eq: gGPEplus}) the typical values for the CdTe experiments conducted in Grenoble \cite{RojanEtAl2017, Maxime}:  $m=4\times10^{-5}$m$_{\textit{e}}$, $\gamma_{l,0}=0.5$ps$^{-1}$, $\gamma_r = \gamma_{l,0}/25$, $g_{\textrm{int}}=7.59 \times 10^{2}$m$\times$s$^{-1}$, $R=4 \times 10^2$m$\times$s$^{-1}$, $P=4\times10^{19}$(m$\times$s)$^{-1}$, $\gamma_{l,2}=1.29$m$^2 \times$s$^{-1}$, $\Omega=100$ THz.
We record the wavefunction $\psi(x,t)$ during the time evolution and extract its phase $\theta(x,t)$ at suitable time intervals for half of the spatial grid, thanks to the symmetry of the potential. 

We work in the low-noise regime, which ensures that the density fluctuations are negligible and also that there are no topological defects, such as solitons or phase slips. This allows us to unwind the phase, $\theta(x,t) \in (-\pi, \pi] \rightarrow \theta(x,t) \in (-\infty, \infty)$. This is necessary since the KPZ universality class describes fluctuations continuously growing in space and time, which cannot be realised for a compact field. In order to achieve the unwinding in the numerical simulations, we constrain the phase such that the difference between neighouring space-time points is less than $0.8 \times 2\pi$, where the factor $0.8$ is chosen empirically in order to take into account unwinding errors due to space and time discretization. We checked that the unwinding protocol is robust, {\it i.e.} the results do not depend on the specific value of this factor as long as it is close to 1.
\begin{figure}[h]
\begin{center}
\includegraphics[scale=0.31]{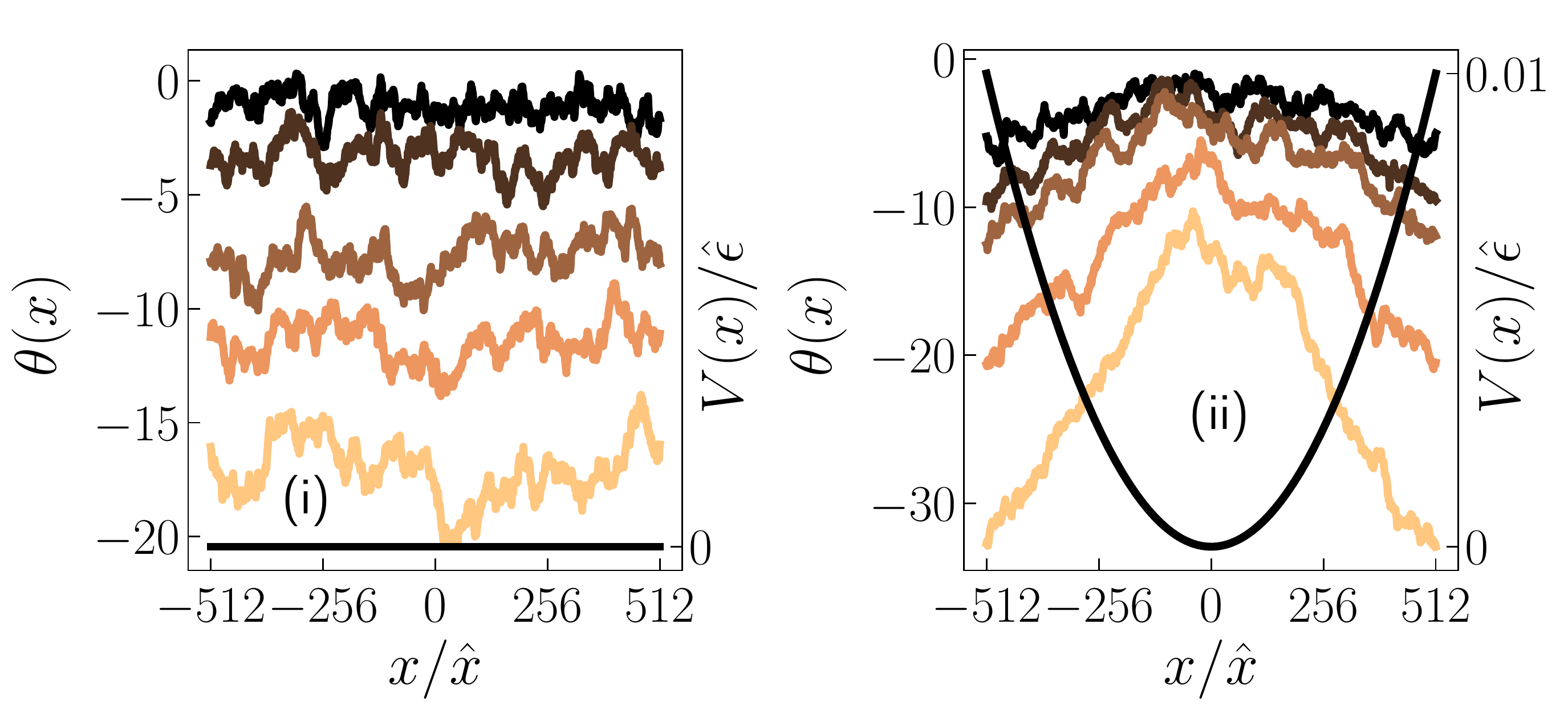}
\caption{Typical spatial phase profiles at different times during the evolution, with lighter colours corresponding to larger times, (i) in the absence of confinement potential, leading to a flat profile, and (ii) in the presence of the parabolic confinement potential, leading to a bent profile.
}
\label{fig: fig1}
\end{center}
\end{figure}

In Fig.~\ref{fig: fig1} we display typical phase profiles obtained in the homogeneous case with no external potential $V=0$, which leads to a flat profile, and in the inhomogeneous case with the parabolic  confinement potential (\ref{eq:pot}), which leads to a curved profile. One can observe that in this case the phase profile indeed propagates faster near the boundaries where the EPs feel the largest potential as anticipated. Around the central tip at $x=0$, the phase presents a local curvature, as we evidence in the following.

\section{Results for the scaling}
The KPZ scaling properties can be studied directly from the first-order correlation function of the EP condensate wavefunction,
\begin{equation}
g_1(\Delta x,\Delta t) =  \dfrac{\left \lvert \left\langle \psi^*(x+\Delta x, t+\Delta t) \psi(x,t) \right \rangle \right \rvert}{\left\langle \sqrt{\rho(x+\Delta x, t+\Delta t) \rho(x,t)} \right\rangle},
\end{equation}
Throughout this work, $\left\langle ... \right\rangle$ denotes the average over noise realisations. Note that the first-order coherence $g_1$ is routinely measured in EP experiments, which renders the following analysis easily accessible. By performing a cumulant expansion and neglecting density-phase correlations, one can relate $g_1$ to the connected correlation function of the phase $C$, obtaining to first order
\begin{align}
-2 \ln \left[ g_1(\Delta x,\Delta t)  \right] & = \left\langle \left[ \theta(x+\Delta x, t+\Delta t) - \theta(x,t) \right]^2 \right\rangle\nonumber\\
 &- \Big\langle \theta(x+\Delta x, t+\Delta t) - \theta(x,t) \Big \rangle^2\nonumber\\
 &\equiv C(\Delta x, \Delta t) \,.
\label{eq: g1}
\end{align}
If the phase follows the 1D KPZ dynamics, it should endow the Family-Vicsek scaling form \cite{FamilyVicsek}
\begin{equation}
 C(\Delta x, \Delta t) = C_0 {\Delta t}^{2/3} g\left(y_0 \frac{\Delta x}{\Delta t^{2/3}}\right)\,,
\label{eq:defC}
\end{equation}
where $g(y)$ is a universal scaling function and $C_0$, $y_0$ are normalisation constants defined as
\begin{equation}
 y_0 = (2A \lambda^2)^{-1/3},\, C_0 =  {\Gamma}^{2/3}, \; A = \frac D \nu, \, \Gamma =\frac{\lambda}{2} A^2\,,
\label{eq:norm}
\end{equation}
where the numerical prefactors are conventional. The precise form of the scaling function $g(y)$ is known exactly only for the stationary interface \cite{Prahofer}. However, the scaling function satisfies the same asymptotics in all sub-classes
\begin{equation}
g(y) \stackrel{y\to 0}{\longrightarrow}  g_0 \; , \quad  g(y) \stackrel{y\to \infty}{\sim} 2 y\,,
\label{eq: gy}
\end{equation}
where $g_0$ is a universal constant depending on the geometrical sub-class, whose values are known exactly \cite{Spohn}.
Therefore, one expects a  similar behaviour for the scaling functions in the three sub-classes, apart from small vertical shifts reflecting the differences in $g_0$ and thus small changes in the intermediate crossover region between the two asymptotic limits. In our simulations, we determined $C(\Delta x,\Delta t)$ from the wavefunction correlation function $g_1$ using Eq.~(\ref{eq: g1}). We first estimated the KPZ scaling exponents with and without the  confinement potential by studying the equal-time and equal-space correlation functions, which, according to Eqs.~(\ref{eq:defC}) and (\ref{eq: gy}), should behave as
\begin{subequations}
\begin{align}
C(\Delta x, \Delta t=0) &\sim  \Delta x^{2\chi}, 
\label{eq: scalinglaw1} \\
C(\Delta x=0, \Delta t) &\sim  \Delta t^{2\beta} \,,
\label{eq: scalinglaw2}
\end{align}
\end{subequations} 
with $\chi=1/2$ and $\beta=1/3$ the 1D KPZ roughness and growth critical exponents. We found $\chi=0.49 \pm 0.01$ and $\beta =0.30\pm 0.01$ in both the flat and the curved cases for the purely spatial and purely temporal correlations, see Fig.~\ref{fig: fig6}. The value of the growth exponent $\beta$ slightly differs from the theoretical one but it is comparable with values reported in previous studies of EP condensate for this system size \cite{Davide} for the flat geometry. 

In order to construct the universal scaling function $g(y)$ defined in Eq.~(\ref{eq:defC}), we first selected all the data points lying in the correct scaling regime by filtering out the points differing by more than small cutoffs $\epsilon_x, \epsilon_t$ from the expected scaling laws in Eqs.~(\ref{eq: scalinglaw1}, \ref{eq: scalinglaw2}) for each value of spatial and temporal separation. We extracted the normalisation parameters $A$ and $\Gamma$ in (\ref{eq:norm}) from our numerical data. Note that in the curved case, these parameters are not homogeneous. However, in the vicinity of the central tip, where the confinement potential is nearly vanishing, they are effectively almost constant and coincide with the values for the homogeneous case (see Sec. \ref{S3} for details and the corresponding data).

\begin{figure}
\begin{center}
\includegraphics[scale=0.31]{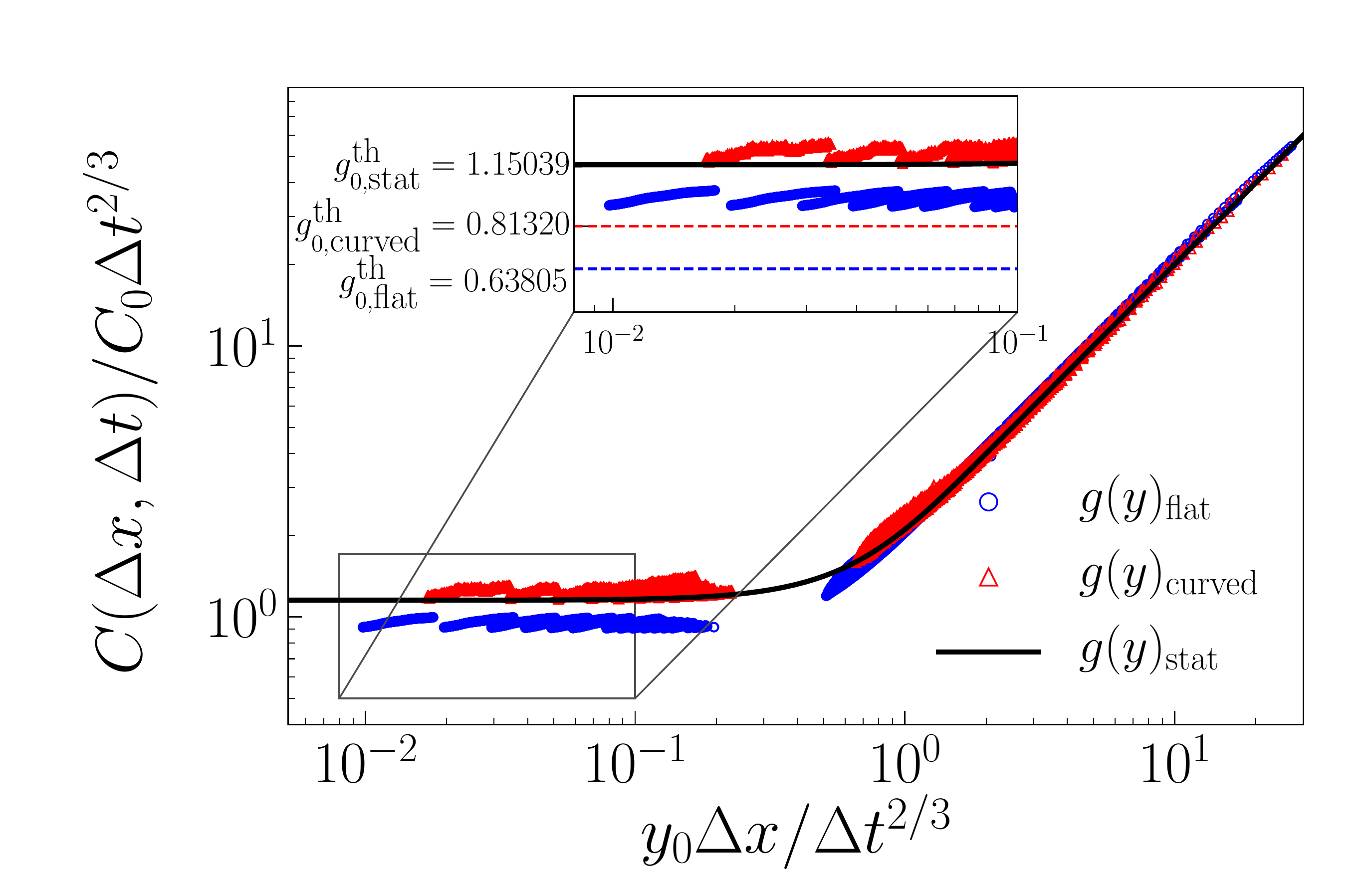}
\caption{Universal scaling function $g(y)$ for the flat (blue dots) and curved (red triangles) phase profiles. The theoretical result for the stationary interface $g_{\textrm{stat}}(y)$ is shown for comparison (solid line). The theoretical values $g^{\rm th}_{0,{\rm flat}}$, $g^{\rm th}_{0,{\rm curved}}$, $g^{\rm th}_{0,{\rm stat}}$ are indicated in the inset, together with the two numerical curves for $y_0 \Delta x/ \Delta t^{2/3} \rightarrow 0$.}
\label{fig: gy}
\end{center}
\end{figure}

The scaling function is obtained by plotting $C(\Delta x,\Delta t)/(C_0 {\Delta t}^{2/3})$ as a function of $y_0 \Delta x/\Delta t^{2/3}$. The results are displayed in Fig.~\ref{fig: gy} together with the theoretical curve $g(y)_{\rm stat}$ for the stationary case \cite{Prahofer}. For both the flat and the curved cases, we observe a reasonable collapse of all the data points onto a single function $g$, which demonstrates that $C$ indeed takes a scaling form. Additionally, we confirm that the scaling functions $g$ are quite similar for the three cases. However, it is not a perfectly one-dimensional curve, it has a finite (small) thickness, and the numerical values for $g_0=g(0)$ differ of about $\sim 40\%$  from the theoretical exact values in both the flat and curved cases  ($g^{\rm num}_{0,{\rm flat}}\simeq  0.95$ vs $g^{\rm th}_{0,{\rm flat}}=  0.63805...$, $g^{\rm num}_{0,{\rm curved}}\simeq  1.23$ vs $g^{\rm th}_{0,{\rm curved}}=  0.8132...$). These discrepancies  may originate from the fact that the actual growth exponent is slightly smaller than the theoretical one, and also from the fact that the $g_1$ function includes other contributions beside the phase correlations, even if they are assumed to be small (higher-order cumulants of the phase or density-phase correlations). However, let us emphasise that the values for $g_0$ are clearly distinct in the two cases, and their ratio (or relative difference) turns out to be within 3\% accuracy with the theoretical ratio (or relative difference). This already indicates that the mapping from the Gross-Pitaevskii to the KPZ equation is well-grounded, and that both the flat and the curved universality sub-classes can be probed in EP systems.

\section{Results for the phase fluctuations}
\subsection{One-point statistics -- Tracy-Widom distributions}
The precise geometry of the phase profile affects the distribution of the fluctuations of the phase. More precisely, as the phase profile propagates linearly in time with fluctuations growing as $t^{1/3}$, one introduces the rescaled fluctuation field $\chi$, defined from the long time behaviour of the phase in 1D as 
\begin{equation}
\theta(x,t) \stackrel{t\to \infty}{\sim} \omega_{\infty}t + (\Gamma t)^{1/3} \chi(\zeta, t)\,,
\label{eq: ansatz}
\end{equation}
where $\omega_{\infty}$ is the asymptotic velocity of the phase, which has a non-trivial dependence on the KPZ parameters \cite{DavideKonstantinosLeonieAnna}, and $\zeta$ is the spatial coordinate rescaled by the correlation length of fluctuations $\zeta \equiv x/\xi(t)$ with $\xi(t)=(\Gamma t)^{2/3}\frac{2}{A}$ \cite{FukaiTakeuchi}. We focus on the universal statistical properties of the centered unwound phase $\Delta \theta(x_0,t) = \theta(x_0,t) - \left\langle\theta(x_0,t) \right\rangle$. 
This allows one to subtract the drift term in Eq.~(\ref{eq: ansatz}), and we henceforth omit the arguments, since there is translational invariance in time all through the KPZ regime (which occurs for stationary condensates), and it is sufficient to consider only the central point $x_0 = 0$. Let us emphasise that Eq.~(\ref{eq: ansatz}) merely stands as an ansatz for the long-time behaviour of the phase. We estimated the numerical value of $\Gamma$ from the time dependence of $\Delta\theta^2$ (see Sec. \ref{S5} for details).

\begin{figure}[h!]
\begin{center}
\includegraphics[scale=0.31]{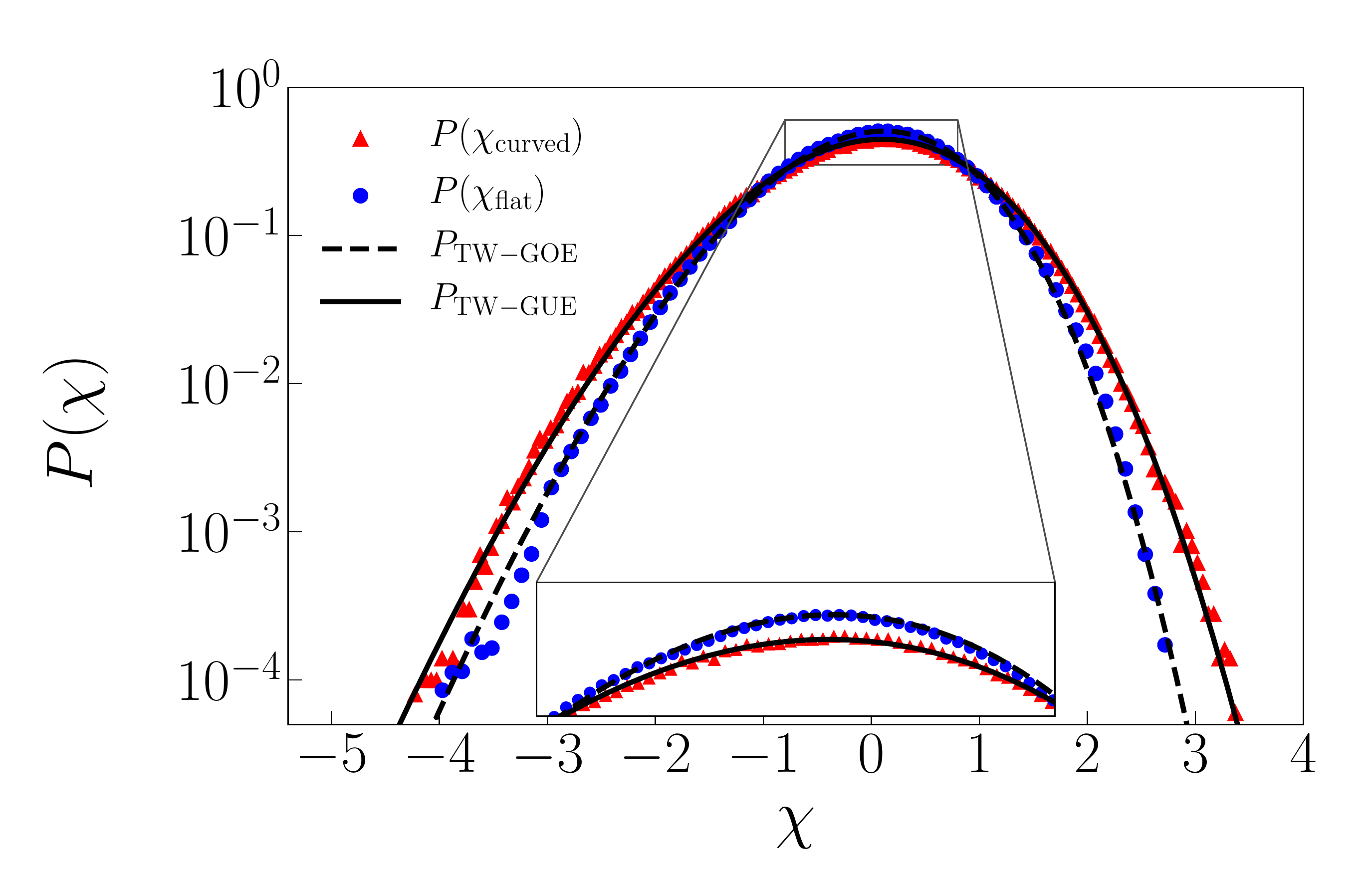}
\caption{Centered distribution of the rescaled phase fluctuations $\chi$ sampled at $x=0$, for flat (blue dots) and curved (red triangles) phase profiles, together with the theoretical TW-GOE and TW-GUE distributions.}
\label{fig: fig3}
\end{center}
\end{figure}

We compute the rescaled fluctuation field $\chi$ from $\Delta\theta$ following  Eq.~(\ref{eq: ansatz}). Note that for the flat case, we conform to the standard definition of the TW-GOE random variable found in the literature, and further rescale $\chi$ as $\chi\to 2^{-2/3} \chi$. We construct the histograms of $\chi$ both for the flat case without confinement and for the curved case with the parabolic confinement $V$. The resulting distributions are displayed in Fig.~\ref{fig: fig3}, where they are compared with the theoretical distributions (more precisely to the mirror ones $P_{\text{TW-GOE}}(-\chi)$ and $P_{\text{TW-GUE}}(-\chi)$ since the KPZ non-linearity $\lambda$ is negative for our choice of experimental parameters).
One observes a clear distinction between the two cases, and moreover the distribution for the curved case is in excellent agreement with the TW-GUE distribution, thus demonstrating that one can indeed tune the KPZ geometrical sub-class realised in the EP system. We emphasise that this result is very robust with respect to the choice of confinement potential, and we found that the agreement remains excellent for a Gaussian walls potential (see Sec. \ref{S4}).

\subsection{Two-point statistics -- correlations of Airy processes}
Besides the probability distribution, the geometry also influences the two-point statistics of the rescaled fluctuations, differing in the three universality sub-classes. In particular, it was shown that the connected correlation function of rescaled fluctuations of the height of the interface at equal time, defined as
\begin{equation} 
C_\chi(\Delta \zeta) \equiv \left\langle \chi(\zeta+\Delta \zeta,t) \chi(\zeta,t) \right\rangle - \left\langle \chi(\zeta+\Delta \zeta,t\right\rangle \left\langle \chi(\zeta,t) \right\rangle\,,
\label{eq:Cchi}
\end{equation}
with $\Delta\zeta=\Delta x / \xi(t)$, is given by the time correlation function of the Airy$_1$, respectively Airy$_2$ process in the asymptotic limit, in the flat, respectively curved geometry 
\begin{equation}
C_\chi(\Delta \zeta) = {\cal G}_i(\Delta\zeta) = \left \langle \mathcal{A}_i(t^{\prime}+\Delta \zeta) \mathcal{A}_i(t^{\prime}) \right\rangle\,,
\end{equation}
where $i=1,2$ stands for Airy$_1$, respectively Airy$_2$, processes. Let us mention one subtle issue here. While the Airy$_2$ process was found to coincide with the dynamics of the largest eigenvalue of GUE matrices \cite{Johansson2}, the Airy$_1$ process differs from the largest-eigenvalue dynamics of GOE matrices \cite{Bornemann2}. This indicates that while one-point statistics of the fluctuations of the growing interface are connected to random matrix theory, this connection is flawed at the two-point level in the flat case. 

\begin{figure}[t!]
\begin{center}
\includegraphics[scale=0.31]{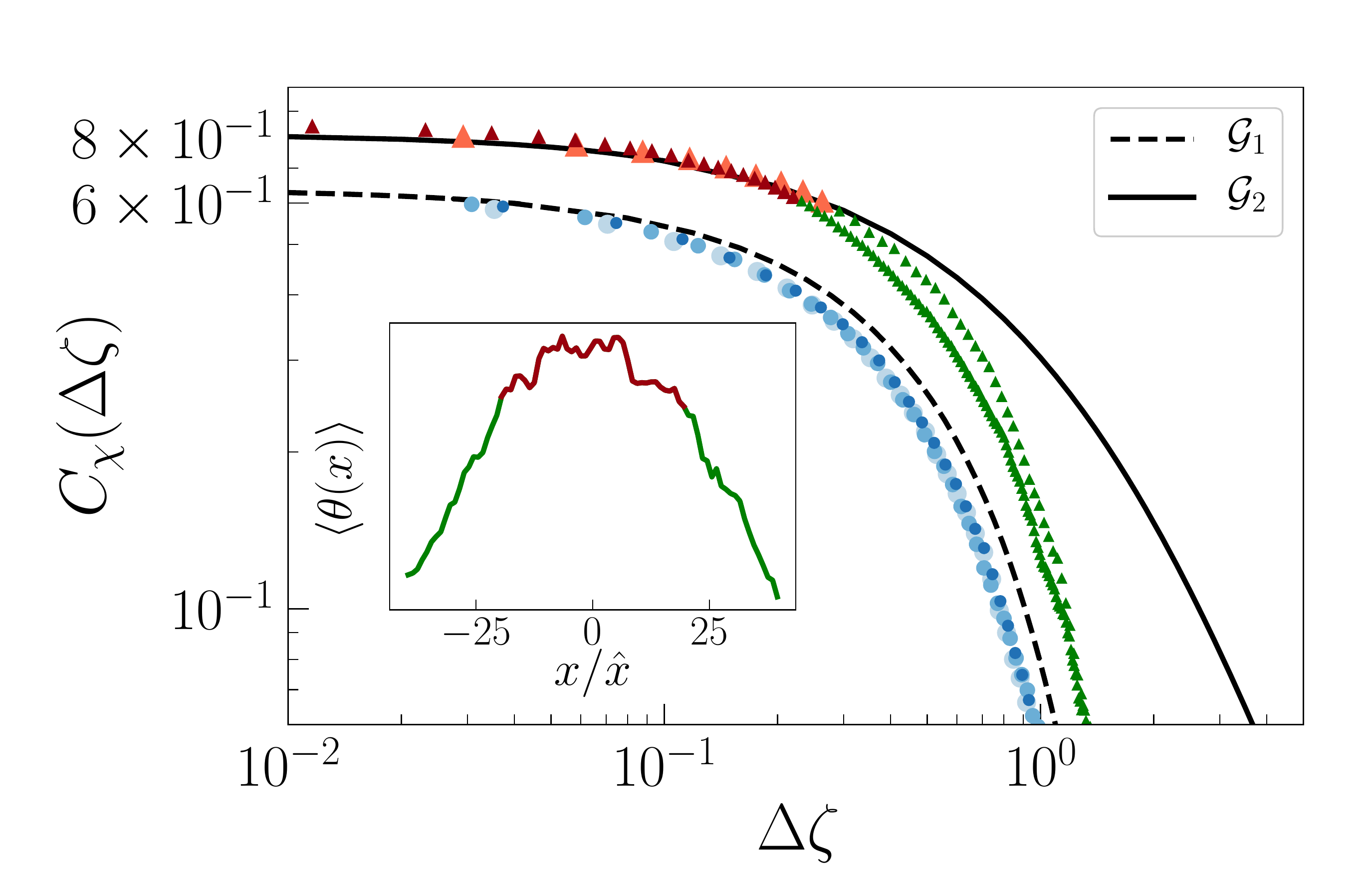}
\caption{Correlation function $C_\chi$ of the rescaled phase fluctuations as a function of the rescaled length $\Delta \zeta$ for different times for the flat case (blue dots) and for the curved case (green and orange triangles), together with the theoretical results for the correlation of the Airy$_1$ ${\cal G}_1(\Delta \zeta)$ (dashed line) and Airy$_2$ ${\cal G}_2(\Delta \zeta)$ (solid line) processes presented in \cite{Bornemann1, Bornemann2, Bornemann3}. For the flat phase profile, data corresponding to  times $t/\hat{t}=7.5\times10^3, 8.125\times10^3, 1\times10^4$ is shown, and for the curved phase profile, data corresponding to $t/\hat{t}=4\times10^2, 1.6 \times 10^3$ is shown (lighter to darker colors correspond to increasing times). For the curved case, the red shades correspond to the small $\Delta \zeta$ regime where the phase profile is locally curved, while green correspond to large $\Delta \zeta$, where the profile is shaped by the effective drag -- see inset.}
\label{fig: fig4}
\end{center}
\end{figure}

In our simulations, we computed the correlation function Eq.~(\ref{eq:Cchi}) of the rescaled phase fluctuations  as
\begin{equation}
C_\chi(\Delta \zeta) = \dfrac{\left\langle \Delta \theta(x,t) \Delta \theta(x+\Delta x, t) \right\rangle}{\left( \Gamma t \right)^{2/3}}\,.
\label{eq: Cchiplot}
\end{equation}
The results we obtained for the two geometries are presented in Fig.~\ref{fig: fig4}. We note that in line with the rescaling of $\chi$ in the flat case mentioned previously, we also perform in this case the following rescaling $t \rightarrow t/2^{-2/3}$ and ${\cal G}_1 \rightarrow {\cal G}_1 / 2^{-2/3}$ to conform to the standard definition of the Airy$_1$ process \cite{TakeuchiSano1} and compare with the theoretical results from \cite{Bornemann1}. Furthermore, in the curved geometry, the limit stochastic process is shown to be $\chi_{\textrm{curved}}(\zeta,t) \stackrel{d}{\rightarrow} \mathcal{A}_2 - \zeta^2$, where $-\zeta^2$ reflects the influence of the mean profile \cite{PrahoferSpohn_2pt1}, which is automatically subtracted in our case since we consider the connected function.

For the flat case, we observe that the correlation function is stable in time, from $t/\hat{t}=7\times10^3$ to, approximately, $t/\hat{t}=1\times10^4$ and we find a good agreement with the theoretical Airy$_1$ correlations ${\cal G}_1(\Delta \zeta)$, even for large $\zeta$. The small shift visible in the figure can be traced back to the fact that the parameter $\Gamma$ in Eq.~(\ref{eq: Cchiplot}) is extracted from the numerical simulations and is found to be a bit larger than the theoretical value (see Sec. \ref{S5}).

For the curved case, the correlation function is stable over a time window from $t/\hat{t}=4 \times 10^2$ until $t/\hat{t} = 1.6 \times 10^3$ where the KPZ scaling regime is observed. We focus on the small$-\Delta \zeta$ limit, corresponding to small spatial separation around $x=0$. Indeed, we expect a local curvature only around the central tip, since the profile near the boundaries is affected by the effective drag ensuing from the confinement potential. The phase correlations are found to behave as an Airy$_2$ process only over this limited range of space, but the agreement with the theoretical Airy$_2$ correlations ${\cal G}_2(\Delta \zeta)$ on this range is extremely satisfactory. 

We emphasise that we have provided in this section the first analysis for both the flat and the curved phase profiles of a very fine statistical quantity: We have here probed reduced correlations, which are  the sub-leading behaviour emerging once the dominant scaling one (studied in the scaling part) has been subtracted. In this respect, the agreement found with the theoretical results for the Airy processes is remarkable. All the results presented here very deeply root in the relevance of KPZ dynamics for the EP system.

\section{Summary and Outlook}
In this work, we have shown that by engineering the confinement potential of 1D exciton polaritons one can tune the geometry of the phase of the condensate and thus access both the flat and curved KPZ universality sub-classes. In particular, we have found excellent agreement with the theoretical exact results, not only for the scaling properties, but also  at the level of one-point statistics (probability distributions), as well as two-point statistics of rescaled fluctuations, though locally for the curved case since the condensate phase is only curved on a limited range by the confinement potential. Our results show the remarkable emergence of KPZ universal properties from the microscopic Gross-Pitaevskii equation for exciton polaritons  at all levels explored so far: not only in the scaling function, but also in the probability distributions and even in the sub-leading behaviour for the two-point correlations, and all this for both the universality sub-classes. We believe that our findings pave the way for stimulating new protocols for investigating KPZ universality in experiments, since the KPZ sub-classes can be accessed through simple engineering of the EP system. In particular, harmonic confinement can be implemented by a suitable engineering of the pumping mechanism \cite{TosiEtAl2012}.

Whereas probing the scaling properties is readily accessible from the measurement of the first-order correlation function, the experimental determination of the probability distributions may be more challenging as it involves the measurement of the time and space resolved phase of the condensate. This requires the development of specific interferometric techniques capable of resolving very small times. On the other hand, higher-order correlations have been measured e.g in \cite{last1,last2} for the case of ultracold atoms. Similar techniques could be implemented in the EP system and lead to the possibility of accessing universal ratios of cumulants, thus enabling the demonstration of the typical non-gaussian shape of the TW-GOE and TW-GUE probability distributions and the characterization of universality sub-classes. Last but not least, the investigation of KPZ universality in 2D is an exciting perspective, both from a theoretical viewpoint, where few indications of KPZ scaling in EP systems have been reported \cite{Zamora2017,Mei2020}, and from an experimental viewpoint, where a high-precision platform for exploring KPZ in 2D is still missing.

\acknowledgments
We acknowledge stimulating discussions with Alberto Amo, Jacqueline Bloch, and  Maxime Richard.  We wish to thank Prof. F. Bornemann for kindly providing us with the theoretical data for the correlation of the Airy processes used in Fig.~\ref{fig: fig4}. K.D. acknowledges the European Union Horizon 2020 research and innovation programme under the Marie Sk\l{}odowska-Curie grant agreement No 754303. L.C. acknowledges support from the French ANR through the project NeqFluids (grant ANR-18-CE92-0019) and support from Institut Universitaire de France (IUF).

\label{Bibliography}
\bibliography{Bibliography}
\appendix

\begin{widetext}

\section{Supplemental Material}
We provide here some complements to the analysis reported in the main text. We first give the detailed derivation of the KPZ mapping in the presence of the confinement potential and discuss its range of validity. We then explore the  influence of the precise shape of the confinement potential on the properties of the curved phase. We also explain in details the determination of the normalisation parameters in both geometries.

\subsection{Dimensionless model}
\label{S1}
We consider the generalised stochastic Gross-Pitaevskii equation for the condensate wavefunction $\psi \equiv \psi(x,t)$ given in the main text:
\begin{align}
i \hbar \partial_t \psi &= \left[\mathcal{F}^{-1}[E_{LP}(k)](x) +V(x) + \hbar g_{\textrm{int}}\left\lvert \psi\right\rvert^2 \right] \psi \nonumber \\
& +\frac{i\hbar}{2}\left[\frac{PR}{\gamma_r}-\frac{PR^2}{\gamma_r^2} \left\lvert \psi \right\rvert^2 - \mathcal{F}^{-1}[{\gamma}_l(k)] \right] \psi + \hbar\xi(x,t) \,,
\label{eq: eq1}
\end{align}
with a quartic approximation of the dispersion of the lower-polariton branch $E_{LP}(k)=\hbar \omega_{0,LP}+\frac{\hbar^2}{2m}k^2 - \frac{1}{2 \hbar \Omega} \left(\frac{\hbar^2}{2m}\right)^2 k^4$, a momentum-dependent polariton loss-rate ${\gamma_l}(k)=\gamma_{l,0} + k^2 \gamma_{l,2}$ and a complex white Gaussian noise of covariance 
$\left\langle \xi(x,t) \xi^*(x^\prime, t^\prime) \right\rangle = 2 \sigma \delta(x-x^\prime) \delta(t-t^\prime)$ with $\sigma = \gamma_{l,0}(p+1)/2$. We recast Eq.~(\ref{eq: eq1}) in dimensionless form by choosing as suitable characteristic time, length and energy scales of the system $\hat{t}=\gamma_{l,0}^{-1}$, $\hat{x}=\sqrt{\hbar/(2m\gamma_{l,0})}$, $\hat{\epsilon} = \hbar \gamma_{l,0}$. This yields
\begin{equation}
i \partial_{\tilde{t}} \tilde{\psi} = \left[-(K_c-iK_d)\partial_{\tilde{x}}^2 - K_c^{(2)} \partial_{\tilde{x}}^4 - (r_c(\tilde{x})-ir_d) + (u_c-iu_d)\left\lvert \tilde{\psi} \right\rvert^2 \right] \tilde{\psi} + \sqrt{\bar{\sigma}} \tilde{\xi}\,,
\label{eq: dimensionlessgpe}
\end{equation}
where all the parameters are related to the microscopic ones of Eq.~(\ref{eq: eq1}) via $r_d=\frac{p-1}{2}, u_c=\frac{g_{\textrm{int}}}{\gamma_{l,0}} \hat{x}^{-1}, u_d=p\frac{R}{2\gamma_r}\hat{x}^{-1},  r_c(\tilde{x})=-\frac{\hbar\omega_{0,LP}+V(\tilde{x})}{\hbar \gamma_{l,o}}, \bar{\sigma} = \sigma \hat t, K_c=1, K_d=\frac{m\gamma_{l,2}}{\hbar}, K_c^{(2)}=\frac{\gamma_{l,0}}{2\Omega}$, with $p$ a dimensionless pumping parameter $p=P/P_{th}=\frac{PR}{\gamma_{l,0} \gamma_r}$.
Henceforth we  omit the tildes, as we work exclusively in dimensionless units.

\subsection{Mapping to the inhomogeneous KPZ equation}
\label{S2}
Starting from Eq.~(\ref{eq: dimensionlessgpe}), we express the wavefunction in the density-phase representation $\psi = \sqrt{\rho} e^{i\theta}$.
We further decompose the density and phase fields as
\begin{equation}
\rho(x,t) = \rho_0(x,t) + \delta \rho(x,t)\, ,\quad\quad\theta(x,t) = \theta_0(x,t) + \delta \theta(x,t)\,,
\end{equation}
where $(\rho_0, \theta_0)$ are defined as the zeroth-order solutions and $(\delta \rho, \delta \theta)$ as small fluctuations around these solutions. 
We obtain $(\rho_0, \theta_0)$ by averaging Eq. (\ref{eq: dimensionlessgpe}) over the noise fluctuations
\begin{subequations}
\begin{align}
& \partial_t \rho_0=-K_d\dfrac{(\partial_x \rho_0)^2}{2\rho_0} + K_d \partial_x^2 \rho_0 - 2\rho_0 K_d(\partial_x \theta_0)^2 - 2K_c (\partial_x\rho_0)(\partial_x \theta_0) - 2\rho_0 K_c \partial_x^2 \theta_0 + 2\rho_0 r_d - 2 \rho_0^2 u_d,\nonumber \\
& \partial_t \theta_0  =- K_c\dfrac{(\partial_x \rho_0)^2}{4\rho_0^2} + K_c\dfrac{ \partial_x^2 \rho_0}{2\rho_0} - K_c(\partial_x \theta_0)^2 + K_d\dfrac{(\partial_x\rho_0)(\partial_x\theta_0)}{\rho_0} + K_d \partial_x^2 \theta_0 + r_c(x) - u_c \rho_0  \,.
\nonumber
\end{align}
\end{subequations}
In the following, we assume that  ${\partial_t \rho_0}\simeq0$, {\it i.e.} that the density reaches a steady state. We further assume that the density fluctuations also attain a stationary state and have a negligible spatial dependence, {\it i.e.} ${\partial_t \delta \rho} \simeq 0$ and $\partial_x^{(n)}\delta \rho \simeq 0$.
We checked in the numerical simulations that this condition is verified in the low noise regime, as illustrated below. The equation for the time evolution of the phase fluctuations $\delta\theta$ then reads 
\begin{align}
\partial_t \delta \theta &= K_d \partial_x^2 \delta \theta - K_c (\partial_x \delta \theta)^2 + \partial_x \delta \theta \left(K_d\dfrac{\partial_x \rho_0}{\rho_0} - 2K_c \partial_x \theta_0 \right) \nonumber\\
& +\delta \rho \left[K_c\dfrac{(\partial_x \rho_0)^2}{2\rho_0^3} - K_c\dfrac{\partial_x^2 \rho_0}{2\rho_0^2} - K_d\dfrac{(\partial_x\rho_0)(\partial_x \theta_0)}{\rho_0^2} - u_c \right] - \sqrt{\dfrac{\bar{\sigma}}{\rho_0}} \text{Re}[{\xi} e^{-\theta_0}]\,,
\label{eq: dtimedeltatheta}
\end{align}
where
\begin{equation}
\delta \rho(x)=\dfrac{2\rho_0 K_d \left[ 2(\partial_x \theta_0) (\partial_x \delta \theta) + (\partial_x \delta \theta)^2 \right] + 2K_c \left[(\partial_x \rho_0)(\partial_x \delta \theta) + \rho_0 \partial_x^2 \delta \theta\right] - 2\sqrt{\bar{\sigma} \rho_0} \text{Im}[{\xi} e^{-i \theta_0}]}{2r_d -4\rho_0 u_d + K_d\frac{(\partial_x \rho_0)^2}{2\rho_0^2} - 2K_d(\partial_x \theta_0)^2 - 2K_c \partial_x^2 \theta_0}\,.
\label{eq: deltarho}
\end{equation}
Note that we used both $\delta \theta \ll \theta_0$ and $ \partial_x^{(n)} \delta \theta \ll \partial_x^{(n)} \theta_0$,
which is  motivated by the fact that the mean phase grows linearly with time, whereas the fluctuations grow with the KPZ exponent $t^{1/3}$. By substituting Eq.~(\ref{eq: deltarho}) into Eq.~(\ref{eq: dtimedeltatheta}) we arrive at an inhomogeneous KPZ equation
\begin{equation}
\partial_t \delta \theta = \nu(x) \partial_x^2 \delta \theta  + \frac{\lambda(x)}{2}(\partial_x \delta \theta)^2 + \sqrt{D(x)} \eta (x,t) + \tilde{v}(x) \partial_x \delta \theta\,,
\label{eq: KPZfinal}
\end{equation}
where the KPZ parameters are given by
\begin{align}
&\nu(x) = K_d + 2\rho_0 K_c \tilde{u}(x), &\lambda(x)=2[-K_c + 2\rho_0 K_d \tilde{u}(x)]\,,
\label{eq: kpzparams1}
\end{align}
and with the presence of an extra term $\tilde{v}(x) \nabla \delta \theta$ that we discuss below, with
\begin{align}
& \tilde{v}(x) =-2K_c \partial_x \theta_0 + 4\rho_0 K_d \tilde{u}(x)\partial_x \theta_0 + K_d\frac{\partial_x \rho_0}{\rho_0} + 2K_c \tilde{u}(x) \partial_x \rho_0 \label{eq: v}\,,
\end{align}
and with the function $\tilde{u}(x)$  defined as
\begin{align}
& \tilde{u}(x) = \dfrac{K_c \left[ \frac{(\partial_x \rho_0)^2}{2\rho_0^3} - \frac{\partial_x^2 \rho_0}{2\rho_0^2} \right]- K_d \frac{(\partial_x \rho_0)(\partial_x \theta_0)}{\rho_0^2} - u_c}{K_d \left[ \frac{(\partial_x \rho_0)^2}{2\rho_0^2} - 2(\partial_x \theta_0)^2 \right] - 2K_c \partial_x^2 \theta_0 + 2r_d -4u_d \rho_0} \label{eq: u} \,.
\end{align}
The noise in Eq. (\ref{eq: dtimedeltatheta}) reads
$$\zeta(x,t)\equiv -2\sqrt{\bar{\sigma} \rho_0} \tilde{u}(x) \text{Im} ({\xi}e^{-i\theta_0}) - \sqrt{\frac{\bar{\sigma}}{\rho_0}} \text{Re} ({\xi}e^{-i\theta_0})\,,$$
which has zero mean, while its strength $D(x)$ can be computed from the covariance $\left\langle \zeta (x,t) \zeta (x^\prime ,t^\prime) \right\rangle$, under the assumption that
\begin{align}
&\left\langle \text{Re}({\xi}(x,t)) \text{Re}({\xi}(x^\prime, t^\prime)) \right\rangle = \left\langle \text{Im}({\xi}(x,t)) \text{Im}({\xi}(x^\prime, t^\prime)) \right\rangle = \delta(x-x^\prime) \delta(t-t^\prime) \nonumber\\
&\left\langle \text{Re}({\xi}(x,t)) \text{Im}({\xi}(x^\prime, t^\prime)) \right\rangle=0 \,.
\end{align}
One finds $\left\langle \zeta (x,t) \zeta (x^\prime ,t^\prime) \right\rangle = 2 D(x)\delta(x-x')\delta(t-t')$
with
\begin{equation}
D(x) = \dfrac{\bar{\sigma}}{2\rho_0} \left(1+4\tilde{u}^2(x) \rho_0^2 \right)\,
\label{eq: kpzparams2}
\end{equation}
and the noise $\eta$ in Eq. (\ref{eq: KPZfinal})
is defined as $\eta = \zeta/\sqrt{D(x)}$.

The confinement potential is {symmetric under $x\rightarrow-x$}, and thus $\theta(x,t)$ and $\rho(x,t)$ are both even functions of $x$. It follows that the function $\tilde v(x)$ is an odd function of $x$ for $\tilde v \partial_x$ to be even (as can be checked on Eq.~(\ref{eq: v})). We are interested in the behaviour of the phase profile around the central tip at $x=0$, since away from it the profile is much affected by the drag at the boundaries and is not expected to follow a KPZ dynamics. Thus, around the tip, $\tilde v(x)\simeq 0$ and this extra term can be neglected in Eq.~(\ref{eq: KPZfinal}), such that one recovers the  KPZ equation with $x$-dependent coefficients given in the main text (after restoring the units).

In order to assess whether the assumptions underlying the previous calculation are fulfilled, we compute the spatial profiles of the zeroth-order solutions for the EP system $(\rho_0, \theta_0)$, the density fluctuations $\delta \rho$, the first spatial derivatives $\partial_x \overline{\rho_0(x)}, \partial_x \overline{\delta \rho(x)}, \partial_x \overline{\theta_0(x)}$ (with the bars denoting time averages) and of the curvature of the phase $\partial_x^2 \theta_0$, in the presence of the shallow harmonic potential introduced in the main text. The zeroth-order solutions are obtained as $\rho_0(x, t) = \left\langle \rho(x,t) \right\rangle$ and $\theta_0(x, t) = \left\langle \theta(x,t) \right\rangle$ in any given $t$. The density fluctuations are then defined as $\delta \rho(x,t) = \rho(x,t) - \rho_0(x)$, and the spatial derivatives are computed numerically. Typical results are displayed in Fig.~\ref{fig: fig1a}.

\begin{figure}[h!]
\begin{center}
\includegraphics[scale=0.4]{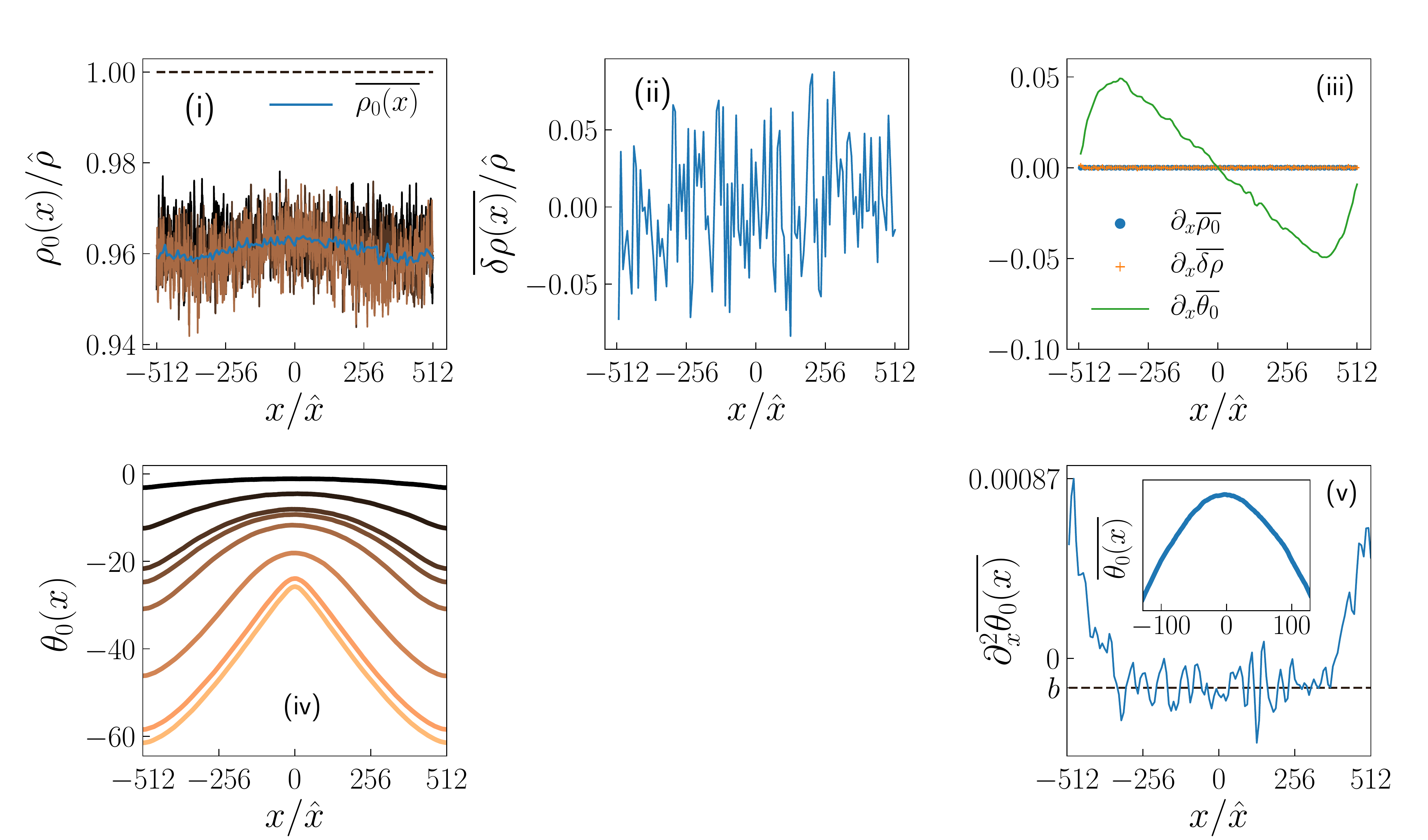}
\caption{(i) Zeroth-order spatial density profile $\rho_0(x)$ for $t/\hat{t}=200, 1400, 2000$ with lighter colours corresponding to larger times, together with the theoretical prediction  $\rho_{\textrm{hom}}=\hat{\rho}{r_d}/{u_d}$ obtained in the homogeneous case for the same parameters and the time average $\overline{\rho_0(x)}$ (blue) in the expected time window for KPZ, which extends approximately from $t/\hat{t}=1400$ until $t/\hat{t}=1600$. (ii) Time averaged spatial density fluctuations $\overline{\delta\rho(x,t)}$ in the same window (iii) and comparison of spatial derivatives $\partial_x \overline{\rho_0(x)}, \partial_x \overline{\delta \rho(x)}, \partial_x \overline{\theta_0(x)}$. For the shallow harmonic potential, $\rho_0(x)$ turns out to be very close to $\rho_{\textrm{hom}}$ and its spatial derivative can be neglected. \\
We further show (iv) the zeroth-order phase profile for $t/\hat{t}=200, 800, 1400, 1600, 2000, 3000, 3800, 4000$, with lighter colours corresponding to later times, demonstrating very slow time dependence in the vicinity of $x=0$ within the KPZ time window, and (v) its curvature after performing a time average in the KPZ window. By fitting the data near $x=0$, we estimate the curvature to be $b \simeq 1.4 \times 10^{-4}$. In the panels (i, ii) the y-axis has been rescaled by $\hat{\rho}=\sqrt{(p-1)\gamma_r/Rp}$ for convenience.}
\label{fig: fig1a}
\end{center}
\end{figure}

We have checked that the time dependence of $\rho_0, \theta_0$ is negligible near $x=0$ for the relevant time window in which KPZ behaviour is observed (see also Fig.~\ref{fig: fig5}), hence it is justified to perform a time average of $\theta_0$ recorded in this window, denoted by the overline bar. The curvature is then found to be a negative constant near $x=0$ as expected. Lastly, $\delta \rho \ll \rho_0$ and $\partial_x \delta\rho \simeq 0$, hence the assumptions are consistent.

Let us note that for such a shallow potential, the spatial variation of the KPZ parameters in Eqs.~(\ref{eq: kpzparams1}), (\ref{eq: kpzparams2}) is small, and in the vicinity of  $x=0$  one finds $\nu(x) \simeq \nu_{\textrm{flat}}, \lambda(x) \simeq \lambda_{\textrm{flat}}, D(x) \simeq D_{\textrm{flat}}$, as shown in Fig.~\ref{fig: fig1b}.
\begin{figure}[h!]
\begin{center}
\includegraphics[scale=0.4]{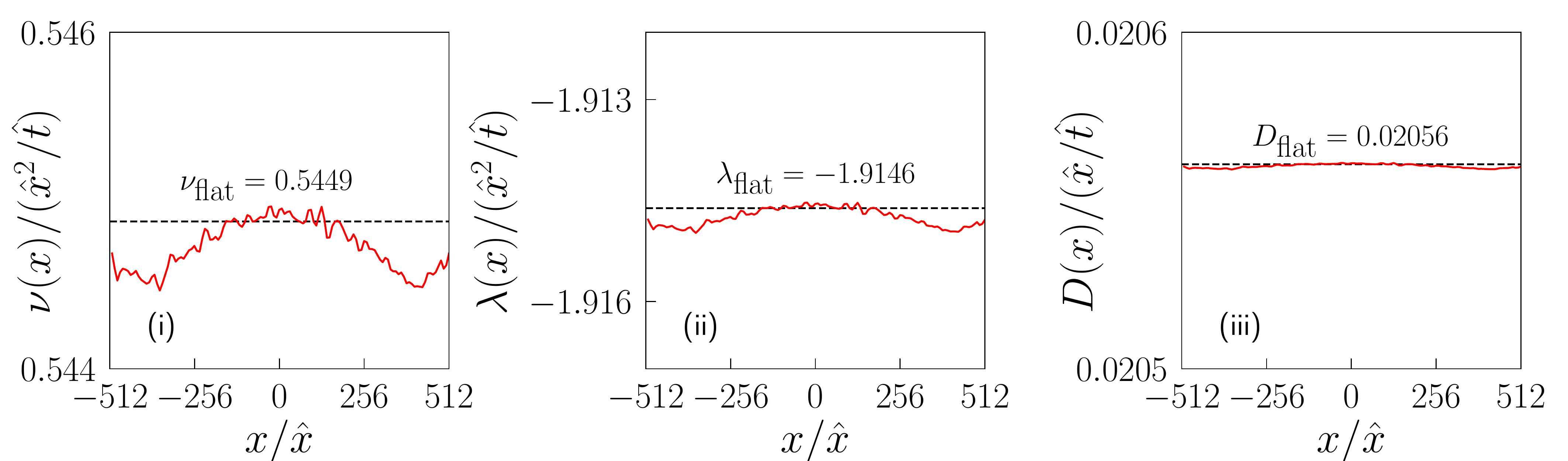}
\caption{KPZ parameters $\nu(x), \lambda(x), D(x)$ obtained after performing the average within the KPZ time window mentioned above, denoted by the overline bar, after restoring the original units. Their spatial variation is very smooth  for our choice of confinement potential and their values in the vicinity of the central point is very close to the homogeneous ones.}
\label{fig: fig1b}
\end{center}
\end{figure}

\subsection{KPZ scaling}
\label{S3}
For completeness, we provide in this section the data for the purely spatial and purely temporal correlation functions calculated from the EP wavefunction first-order correlation $g_1$. 
They are expected to behave as
\begin{subequations}
\begin{align}
C(\Delta x, \Delta t=0) &\sim A \Delta x, \label{eq: scalinglaw11} \\
C(\Delta x=0, \Delta t) &\sim g_0 \Gamma^{2/3} \Delta t^{2/3}\,.
\label{eq: scalinglaw22}
\end{align}
\end{subequations}
Our results are shown in Fig.~\ref{fig: fig6}. 
\begin{figure}[h!]
\begin{center}
\includegraphics[scale=0.27]{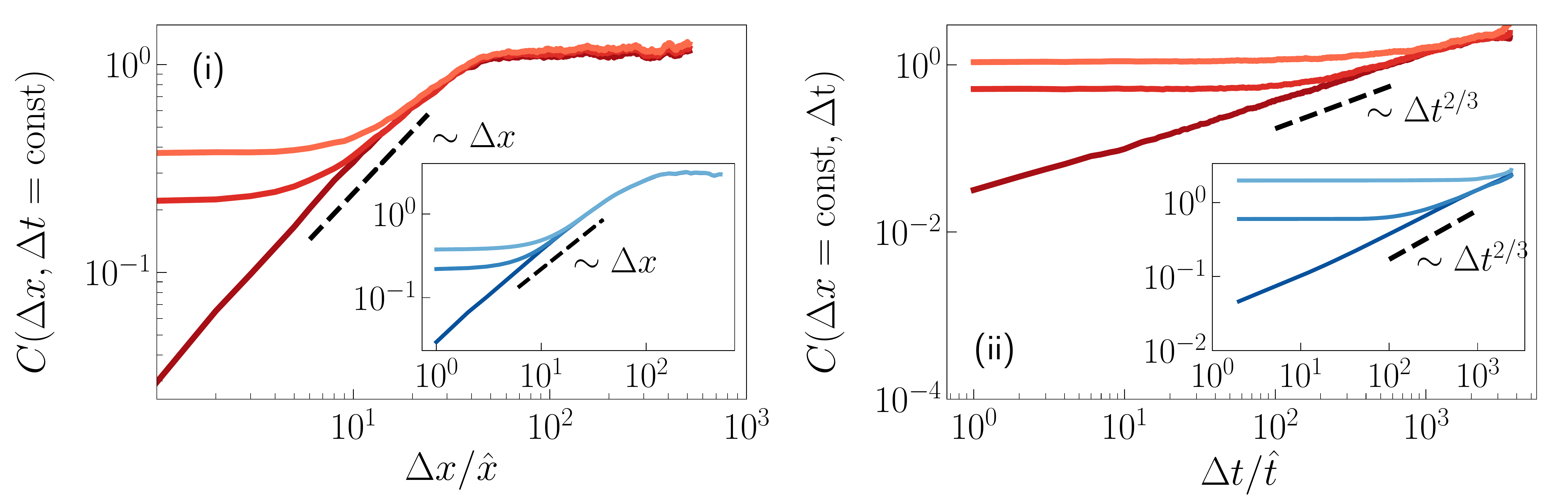}
\caption{(i) Purely spatial and (ii) temporal correlation functions, with the scaling laws as guide to the eye for each case (dashed lines), for the curved (main plots) and flat (insets) geometries. We display three curves which correspond to different fixed values of temporal and spatial separation, $\Delta t /\hat{t}=0,40,100$ and $\Delta x/\hat{x}=0, 16, 64$ (darker to lighter shades).}
\label{fig: fig6}
\end{center}
\end{figure}
We observe a good agreement with the theoretical scaling laws for both the flat and the curved case, with $\chi\simeq 0.49$ and $\beta\simeq 0.30$ compared to the exact values $\chi=1/2$ and $\beta=1/3$.

Furthermore, one can extract the value of the parameter $A$ defined in Eq.~(\ref{eq: scalinglaw11}) from the spatial correlation. We find
\begin{equation}
A \simeq
\begin{cases}
&0.032 \hat x^{-1}, \text{ curved geometry} \\
&0.035\hat x^{-1}, \text{ flat geometry}.
\end{cases}
\end{equation}
These values can be compared with the theoretical one $A_{\textrm{th}} \equiv D/\nu$. For the flat case, $D$ and $\nu$ are related to the microscopic parameters through Eqs.~(\ref{eq: kpzparams1}) and (\ref{eq: kpzparams2}) with $\theta_0=0$ and $\rho_0 = \rho_{\textrm{hom}}=\hat{\rho}r_d/u_d$. One finds $A_{\textrm{th}}\simeq 0.0377\hat x^{-1}$, which is in very close agreement with the values extracted from the spatial correlation. For the curved case, we focus on spatial points around the central tip. As shown in Fig. \ref{fig: fig6},   although the parameters $D$ and $\nu$ depend on $x$, they are nearly constant in the vicinity of $x=0$ and very close to the values for the flat case, hence one finds $A(x)\simeq A(x=0) \simeq A_{\rm flat}$.

\subsection{Influence of the confinement potential}
\label{S4}
\subsubsection{Definition of a Gaussian walls potential}
In order to test the robustness of our proposal to engineer a curved phase profile, we investigate the influence of the precise shape of the confinement potential by studying two different potentials: a harmonic potential which corresponds to the analysis reported in the main text, as well as  Gaussian walls defined as
\begin{equation}
V_{G}(x) = \dfrac{V_{0}}{\ell} \left(e^{-\left(\frac{x-L/2}{\ell}\right)^2} + e^{-\left(\frac{x+L/2}{\ell}\right)^2} \right)\,,
\label{eq: gaussian}
\end{equation}
which interpolate, for a given strength $V_0$, between a hard-wall potential for $\ell\rightarrow 0$ and a smooth well  for $\ell \gg 0$, as illustrated in Fig.~\ref{fig: fig7}. Note that in the simulations, since $\ell$ is expressed in units of the characteristic length $\hat{x}$, $\ell < 1$ effectively corresponds to the $\ell \rightarrow 0$ limit.
\begin{figure}[h!]
\begin{center}
\includegraphics[scale=0.3]{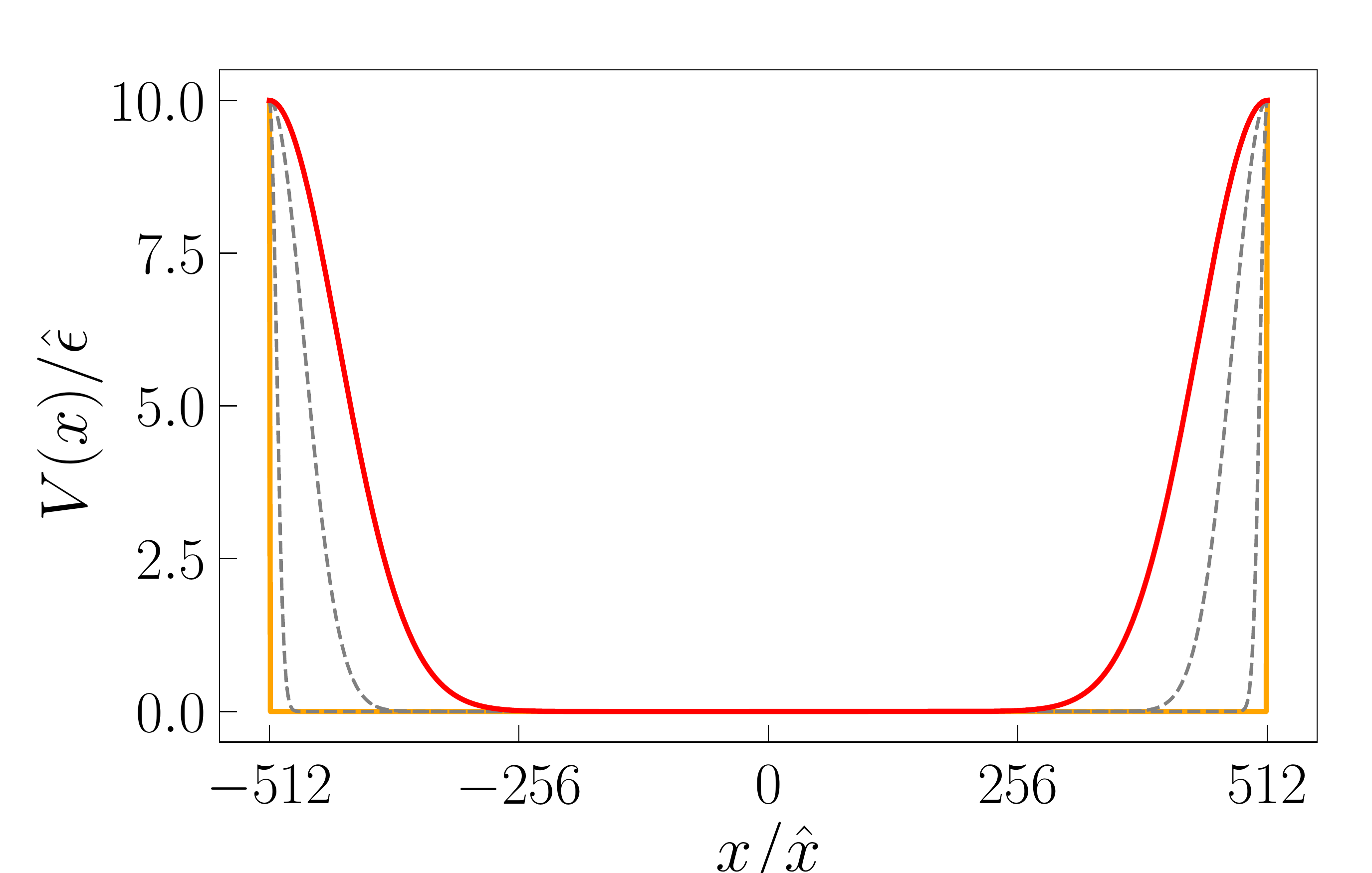}
\caption{Gaussian walls potential for different values of the parameter $\ell=0.01,10,50, 100$, which interpolate between hard walls ($\ell=0.01$, yellow) and smooth Gaussian walls ($\ell=100$, red). The ratio $V_0/\ell=10$ is kept constant.}.
\label{fig: fig7}
\end{center}
\end{figure}
Typical phase profiles obtained for the hard-wall potential ($\ell=0.01$) and the smooth Gaussian walls ($\ell=100$) are shown in Fig.~\ref{fig: fig8}.
The hard-wall potential initially only affects the boundaries and then slowly bends the phase profile. For larger values of $\ell$, a larger portion of the phase profile immediately feels the potential,  and  the bending occurs more rapidly. With this potential, the phase profile at short times  display a smoother curvature around the central tip. We thus focus in the following on the smooth Gaussian walls corresponding to $\ell=100$.
\begin{figure}[t!]
\begin{center}
\includegraphics[scale=0.43]{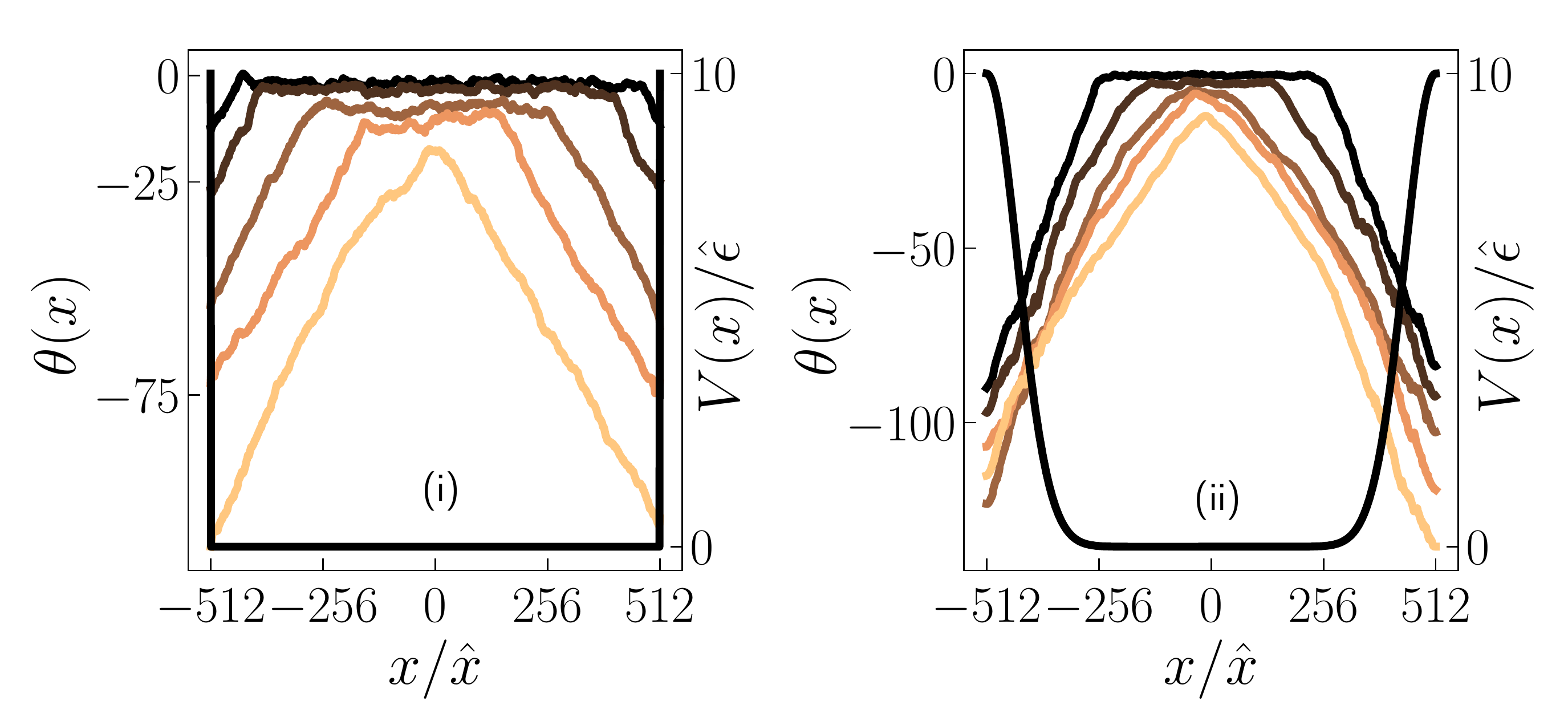}
\caption{Typical phase profiles at different times during the evolution, with lighter colours corresponding to larger times, together with the relevant potentials (black). In the presence of (i) the hard-wall potential ($\ell=0.01$), the phase profile is displayed  for $t/\hat{t}=3\times10^2, 6\times10^2, 1.4\times10^3, 2\times10^3, 2.8\times10^3$ and (ii) the smooth Gaussian walls ($\ell=100$), it is shown for $t/\hat{t}=1\times10^2, 5\times10^2, 1\times10^3, 1.2\times10^3, 1.5\times10^3$ . Note that for the latter we averaged over 6 realizations of the noise in order to stabilize the profile.}
\label{fig: fig8}
\end{center}
\end{figure}
For all values of $\ell$, at late time, a cusp forms at the central tip and the phase is no longer smoothly curved. The KPZ regime is thus expected to develop only at intermediate times.

\subsubsection{Scaling of the variance}
In order to assert the presence of the KPZ regime, we first study the scaling of the variance
of the phase profile {$\langle \Delta \theta (x_0,t)^2\rangle$} at various space points $x_0$, with 
\begin{equation}
\Delta \theta(x=x_0, t) = \theta(x_0, t) - \left\langle \theta(x_0,t) \right\rangle \,,
\end{equation}
and where $\left \langle ... \right \rangle$ denotes the average over noise realisations. This quantity is expected to behave as {$\left\langle \Delta\theta(x_0, t)^2 \right \rangle \sim t^{2/3}$} if the dynamics is in a KPZ regime. Our results are displayed in Fig.~\ref{fig: fig5} for both the smooth Gaussian walls and the harmonic potential. For the Gaussian walls, one observes a KPZ dynamics at short and intermediate time, before a sharp crossover occurs to another regime at long times. The time of the crossover depends on the location on the phase profile. It corresponds at $x=0$ to the time where the cusp is formed, and this is where the KPZ regime is the most extended in time. It is also clear that the KPZ scaling extends over a longer time for the parabolic potential, which is thus more favorable to study the KPZ dynamics. However, we found that even with  Gaussian walls, the KPZ advanced statistics can still be precisely observed, as shown below. We focus in the following on the central tip $x=0$ (for one-point statistics), and hence omit the arguments.
\begin{figure}[h!]
\begin{center}
\includegraphics[scale=0.43]{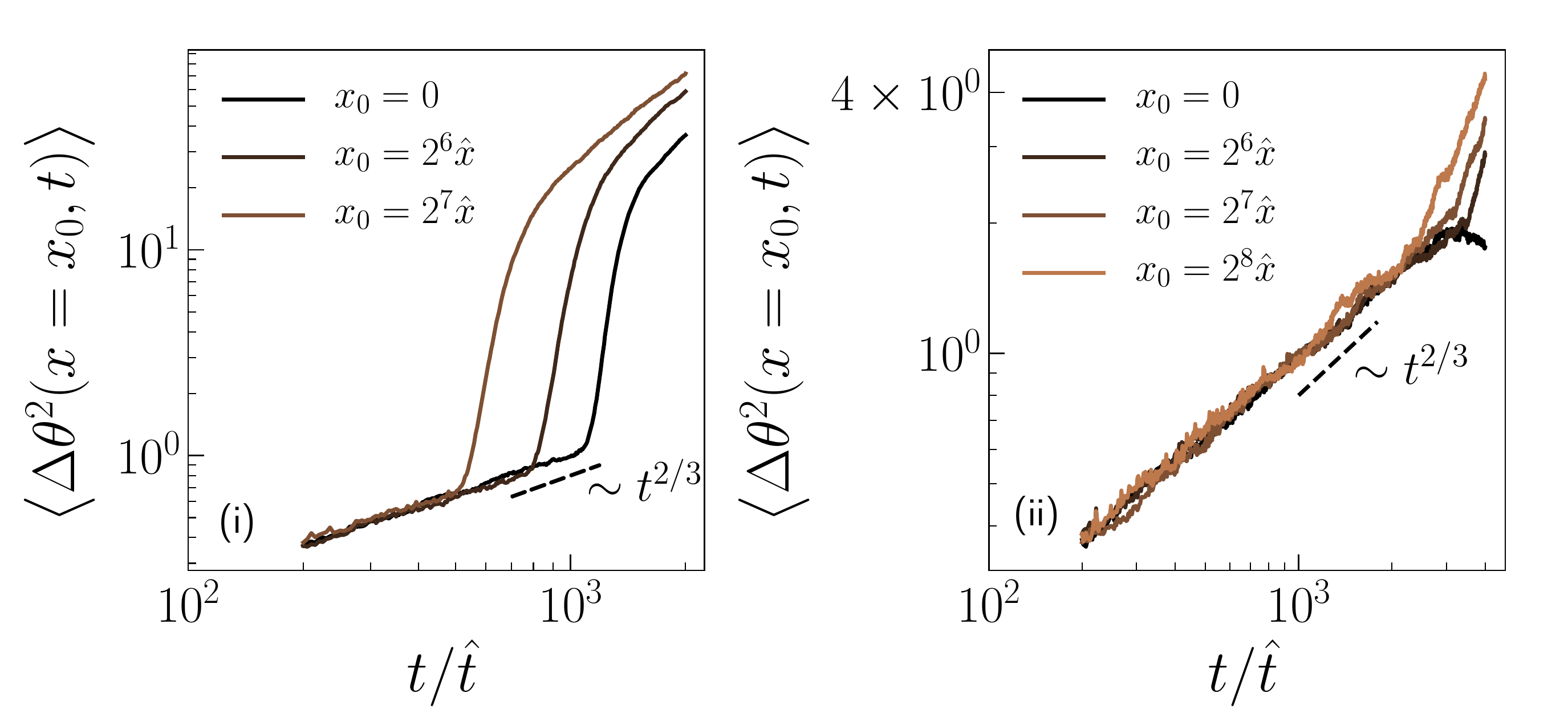}
\caption{Variance of the phase computed at different $x_0$ for the evolution with (i) the smooth Gaussian walls and (ii) the parabolic potential. The KPZ scaling persists for longer times around the central tip for the Gaussian walls, and still longer times for the parabolic potential.}
\label{fig: fig5}
\end{center}
\end{figure}

\subsubsection{Probability distribution and correlation function for the smooth Gaussian walls}
We computed for the smooth Gaussian walls the probability distribution and the correlation of the rescaled height fluctuation $\chi$ defined from the long-time limit of the phase field as
\begin{equation}
\theta \stackrel{t\to \infty}{\sim} \omega_{\infty}t + (\Gamma t)^{1/3} \chi\,.
\label{eq:thetainf}
\end{equation}
The method is the same as the one described in the main text. The results are displayed in Fig.~\ref{fig: fig10} and \ref{fig: fig11}, to be compared with the corresponding Fig. 3 and 4 of the main text obtained for the parabolic potential. One observes that the probability distribution still follows with great accuracy a TW-GUE distribution, as for the parabolic potential, although the time window of the KPZ dynamics is shorter. We also computed the probability distribution of the rescaled phase fluctuations for the hard-wall potential, and the agreement is as remarkable, the two curves being in fact superimposed. For the two-point statistics, the agreement with the theoretical curve for the Airy$_2$ process is still satisfactory for small $\Delta \zeta$. As for the parabolic potential, the phase acquires a smooth curvature only around the central point, such that the spatial region where the universal properties of the KPZ curved sub-class can be observed has a limited extension.

These results show that the proposed protocol to engineer the KPZ universality sub-classes is remarkably robust, since the KPZ statistics related to the curved geometry are found to be nearly insensitive  to the precise form of the confinement potential. The main change between the various confinement potentials is the time and space windows over which the KPZ regime is realised.
\begin{figure}[ht!]
\begin{center}
\includegraphics[scale=0.3]{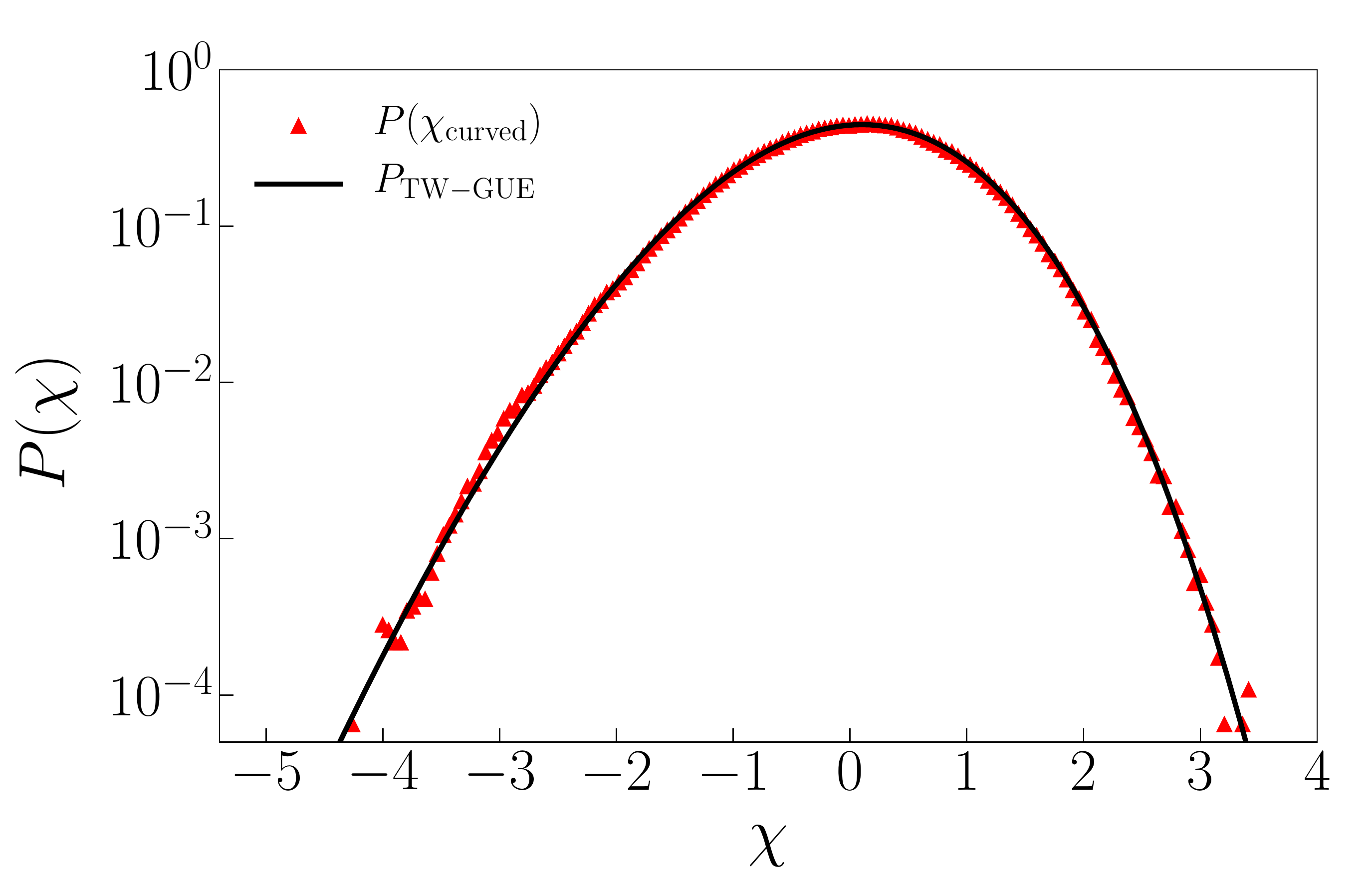}
\caption{Centered distribution of the rescaled phase fluctuations $\chi$ sampled at $x=0$, for the curved phase profile with smooth Gaussian walls, together with the theoretical TW-GUE distribution.}
\label{fig: fig10}
\end{center}
\end{figure} 
\begin{figure}[t!]
\begin{center}
\includegraphics[scale=0.3]{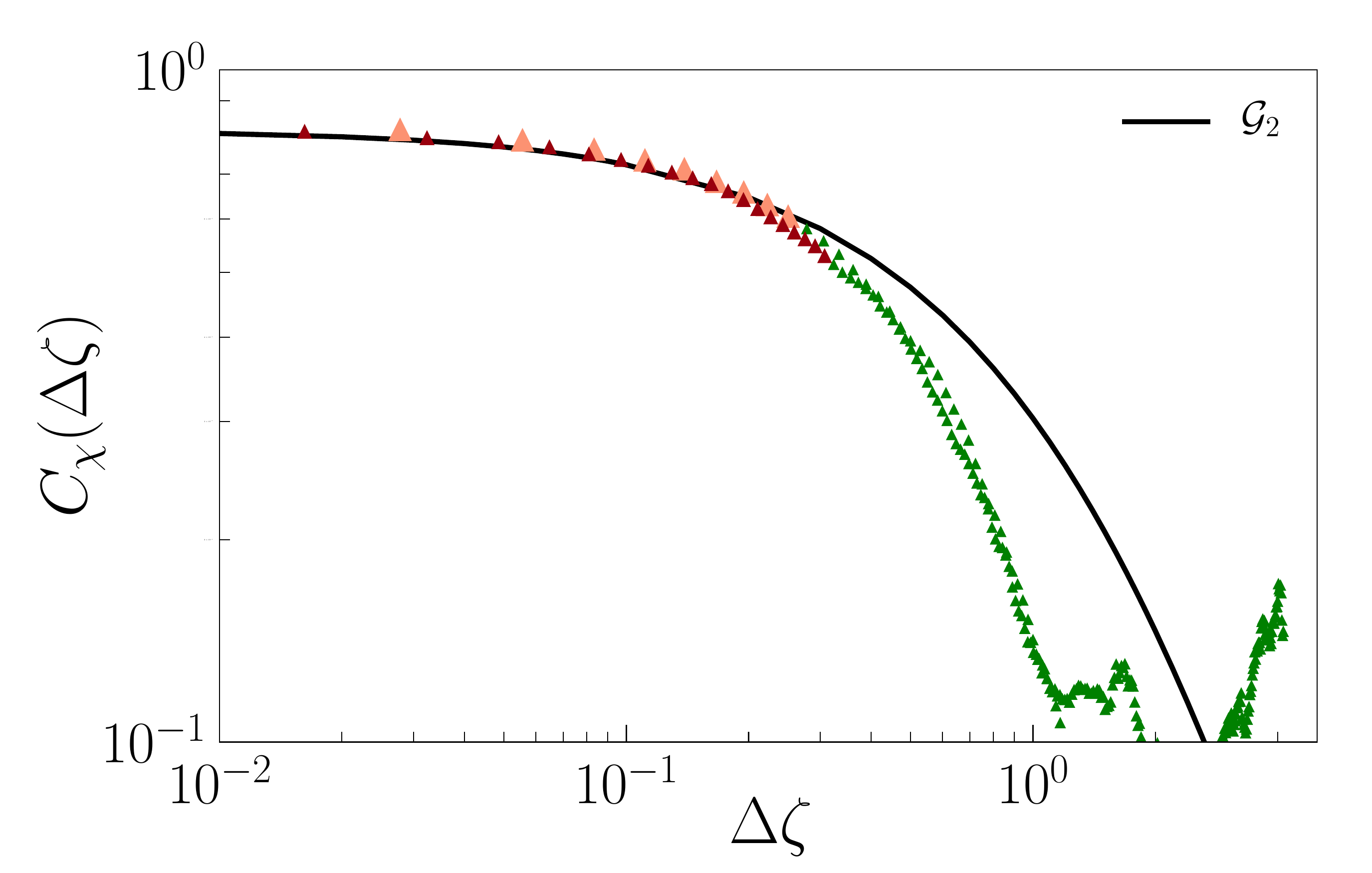}
\caption{Correlation function $C_\chi$ of the rescaled phase fluctuations as a function of the rescaled length $\Delta \zeta$ for the curved phase with the smooth Gaussian walls, together with the theoretical results for the correlation ${\cal G}_2(\Delta \zeta)$ of the  Airy$_2$ process. Data corresponding to times $t/\hat{t}=4\times10^2$ (orange triangles), $9\times10^2$ (red triangles) is shown. The red shades roughly correspond to the small $\Delta \zeta$ regime where the phase profile is locally curved, while green correspond to large $\Delta \zeta$, where the profile is shaped by the effective drag, in the same spirit as for Fig.~4 of the main text.}
\label{fig: fig11}
\end{center}
\end{figure}

\subsection{Numerical estimation of the $\Gamma$ parameter}
\label{S5}
In this work, we computed the scaling function associated with the two-point correlation function of the phase, and the probability distribution and spatial correlation function of the rescaled phase fluctuations $\chi$. The comparison with the theoretical exact results for these quantities requires to fix some normalisations. These normalisations are defined in terms of $A$ and $\Gamma$ defined in Eq. (9) of the main text, which in turn are related to the 
parameters $D$, $\nu$ and $\lambda$ of the KPZ equation (\ref{eq: KPZfinal}). The theoretical expressions of these parameters in terms of the microscopic parameters of the Gross-Pitaevskii equation are established from the mapping to the KPZ equation in Sec. \ref{S2}. For our choice of parameter values (corresponding to Grenoble experiments), we obtain $\Gamma_{\textrm{th}} =\frac{\lambda A^2}{2}\simeq 0.00136 \hat t^{-1}$ and $A_{\textrm{th}}=\frac{D}{\nu}\simeq 0.0377 \hat x^{-1}$. As shown in Sec. \ref{S2}, the numerical values of $A$ extracted from the asymptotic behaviour of the $g_1$ function in the flat and curved cases are very close to the theoretical estimate. 
  
Let us now discuss the numerical values  of $\Gamma$. This parameter can be extracted from our simulations from the definition of the  long-time ansatz Eq.~(\ref{eq:thetainf}). It follows from this definition that the  variance of the phase is related to the variance of the rescaled fluctuations $\chi$ as
\begin{equation}
\left\langle \Delta\theta^2 \right\rangle = (\Gamma t)^{2/3} \text{Var}(\chi)\,,
\end{equation}
where the value of $\text{Var}(\chi)$ is known exactly in both geometries \cite{Spohn}.
In order to extract $\Gamma$, we thus compute  $\left\langle \Delta\theta^2 \right\rangle/t^{2/3}$, average over the plateaus reached in the appropriate time windows corresponding to the KPZ regime and divide by $\text{Var}(\chi)$. The values obtained before the rescaling by the variance are illustrated on Fig.~\ref{fig: fig12} where these plateaus are shown.
\begin{figure}[t!]
\begin{center}
\includegraphics[scale=0.3]{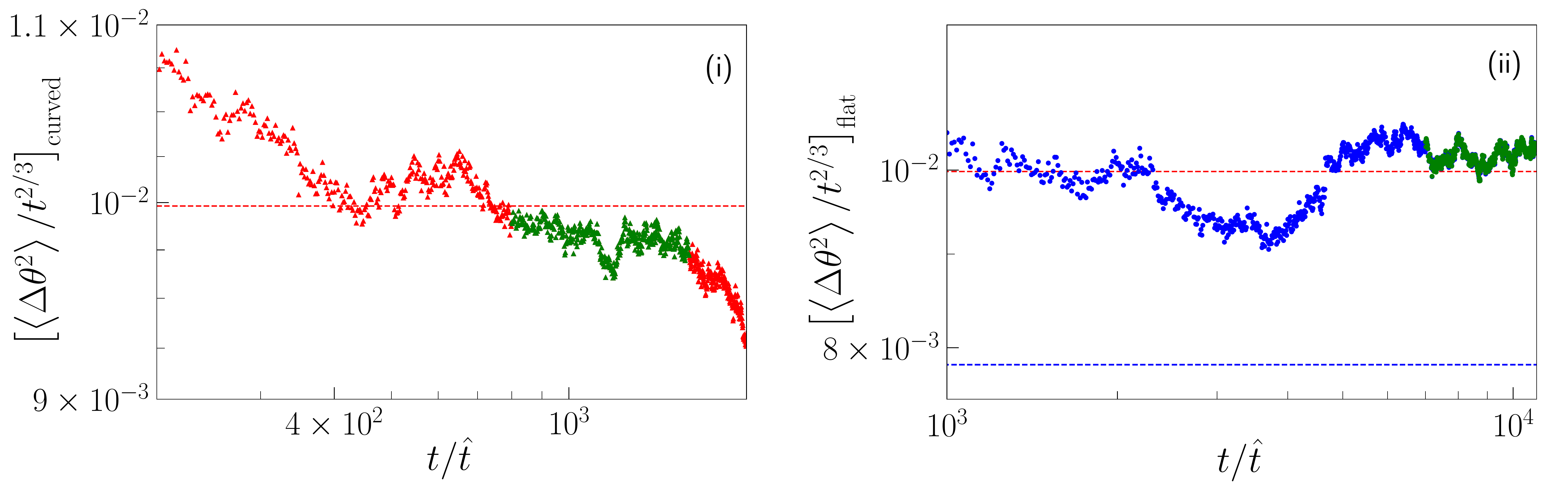}
\caption{Determination of the $\Gamma$ parameter in the (i) curved  and (ii) flat geometry, together with the theoretical values for $\Gamma^{2/3} \text{Var}(\chi)$ (red and blue dashed lines) obtained from the microscopic parameters with the theoretical value of  $\text{Var}(\chi)$ for the identified distribution in each case. The rescaled variance  $ \left\langle \Delta\theta^2 \right\rangle/t^{2/3}$ is averaged over the plateaus in the green time windows and then divided by $\text{Var}(\chi)$, which yields  the values listed in Eq.~(\ref{eq: gammavalues}).}
\label{fig: fig12}
\end{center}
\end{figure}
We obtain
\begin{equation}
\Gamma\simeq
\begin{cases}
&0.0013\hat t^{-1}, \text{ curved geometry} \\
&0.002\hat t^{-1}, \text{ flat geometry}.
\end{cases}
\label{eq: gammavalues}
\end{equation}
These values are again in agreement with  the theoretical estimate $ \Gamma_{\textrm{th}}$. As previously, eventhough $\Gamma$ depends on $x$ in the curved case, it is almost constant if one restricts to a small space region around  the central point $x=0$.  We observe that while the data for the curved phase lies very close to the corresponding theoretical prediction (before the departure from KPZ universality for large times), the value for the flat phase  differs by about 30\% from the theoretical one. We are, at present, unable to explain this small discrepancy, but nevertheless we take it into account in our work by using in the normalisations the actual numerical values of $\Gamma$.

\subsection{Higher-order cumulants}
\label{S6}
To compare the probability distribution of the phase fluctuations and their correlation function to the theoretical ones requires to fix the normalisations, which involve $\Gamma$. To determine  $\Gamma$, one needs to choose the value of Var($\chi$), whether the GOE or the GUE one, and thus to guess a priori which of the sub-class is realised. However, let us emphasise that the sub-class can be determined without any prior knowledge by computing  universal ratios of cumulants, which are independent of the normalisations. We now present this analysis. We compute the third and fourth order cumulants of the centered phase $\left\langle \Delta\theta^3 \right \rangle_c 
\equiv\left\langle \Delta\theta^3 \right \rangle$ and $ \left\langle \Delta\theta^4 \right \rangle_c \equiv \left \langle \Delta \theta^4 \right \rangle - 3 \left \langle \Delta\theta^2 \right \rangle^2$.  From them, one can construct the skewness $\text{sk}(\Delta \theta) = \left \langle \Delta \theta^3 \right \rangle_c / \left \langle \Delta\theta^2 \right \rangle^{3/2}$ and excess kurtosis $\text{ku}(\Delta \theta) = \left \langle \Delta\theta^4 \right \rangle_c / \left \langle\Delta \theta^2 \right \rangle^{2}$, which are universal. They can be compared with the theoretical values for $\text{sk}(\chi)$ and $\text{ku}(\chi)$ which are known exactly for both the GOE and GUE distributions \cite{Spohn} and do not depend on the non-universal parameters $\omega_{\infty}$ or $\Gamma$.
\begin{figure}[h!]
\begin{center}
\includegraphics[scale=0.26]{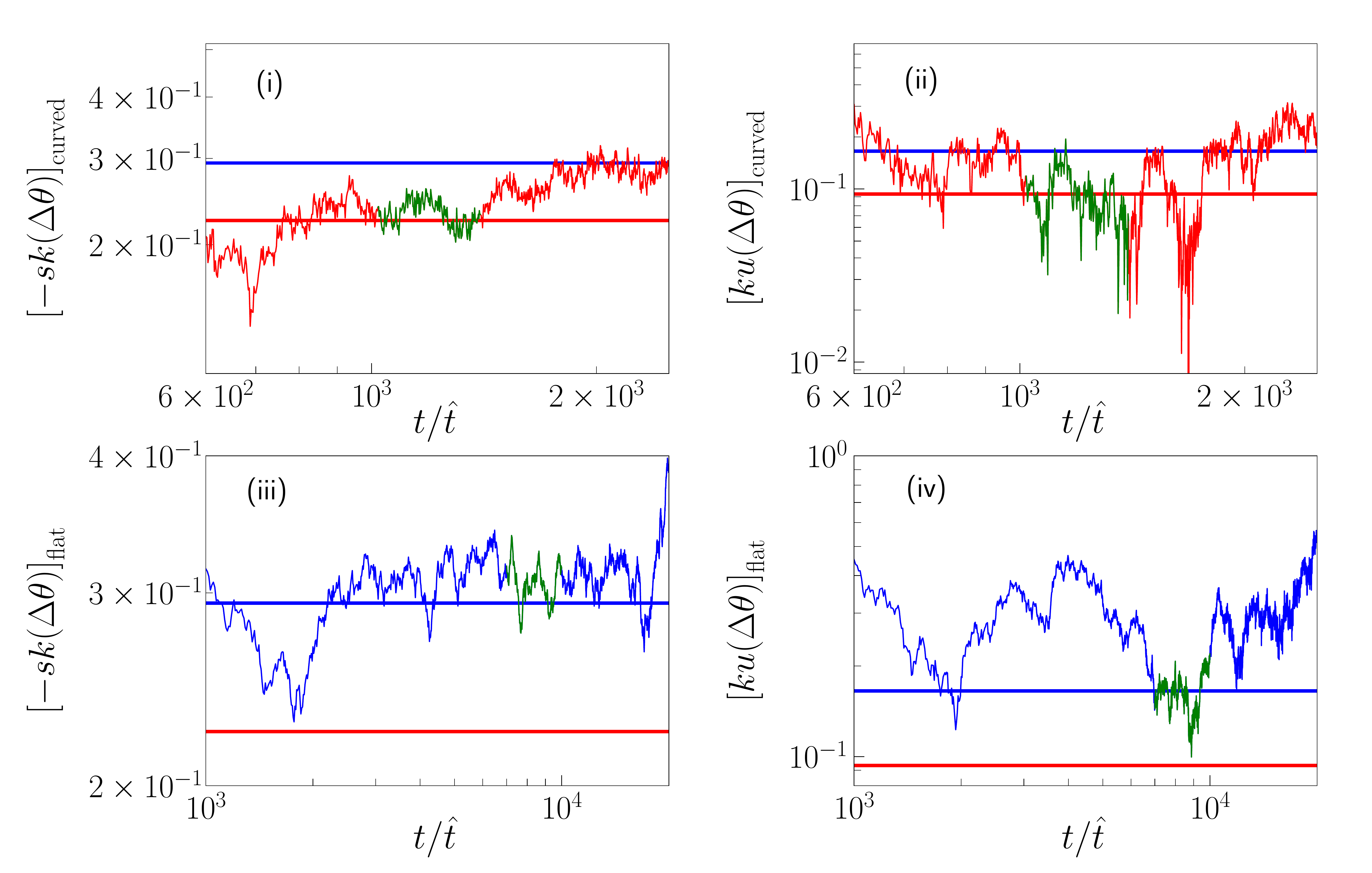}
\caption{Skewness and excess kurtosis of the centered unwound phase field of the condensate in the (i, ii) curved and (iii, iv) flat geometries, together with the theoretical values corresponding to the TW-GOE (blue) and TW-GUE distribution (red). We also display the universal plateaus reached for each geometry (green).}
\label{fig: fig13}
\end{center}
\end{figure} 
Our results are presented in Fig.~\ref{fig: fig13} for both geometries, where the curved one corresponds to evolution under the harmonic potential. Let us note that we rather compute $-sk(\Delta \theta)$ because we are interested in the mirror distribution. 

For the flat geometry, since the profiles are homogeneous, we performed an additional spatial average in order to accumulate statistics. We find convergence to the anticipated values for both cases, in particular for the skewness, in the same time windows as for the variance corresponding to the KPZ regime. For the curved geometry, we observe at large times a departure from the stationary plateaus which correspond to KPZ universality, in line with previous observations. The excess kurtosis, which involves the determination of the fourth-order cumulant, is naturally less statistically tame, even for the case of the curved geometry where averaging over $10.000$ independent realisations of the noise has been performed at the chosen space point $x=0$.

These results are an independent confirmation that the fluctuations of the phase with or without confinement follow two different distributions, and their skewness and excess kurtosis coincide with the ones expected for a curved or flat geometry respectively. This justifies a posteriori the choice of $\text{Var}(\chi$) in the previous section. Let us emphasise that these quantities may be easier to measure experimentally and could be used as a direct probe of the strong non-Gaussianity of the distributions,  and also of the existence of different universality sub-classes for these distributions.

\end{widetext}
\end{document}